\documentclass[twocolumn,showpacs,preprintnumbers,amsmath,amssymb]{revtex4}
\usepackage{graphicx}
\usepackage{dcolumn}
\usepackage{bm}

\usepackage{mathptmx}
\usepackage{multirow}

\newcounter{saveeqn}

\renewcommand{\vec}[1]{{\bf #1}}
\newcommand{\vk}{\vec{k}}
\newcommand{\vq}{\vec{q}}
\newcommand{\vG}{\vec{G}}
\renewcommand{\vr}{\vec{r}}

\newcommand{\tpL}{\tilde{\phi}^{\Lambda}}

\newcommand{\bra}[1]{\langle #1 \,|}
\newcommand{\ket}[1]{|\,#1\rangle}
\newcommand{\braket}[2]{\langle #1 | #2 \rangle}
\begin{document}

\title{Efficient $\mathcal{O}(N^2)$ approach to solve the Bethe-Salpeter equation for excitonic bound states}

\author{F. Fuchs, C. R\"{o}dl, A. Schleife, and F. Bechstedt}
\affiliation{Institut f\"ur Festk\"orpertheorie und -optik, Friedrich-Schiller-Universit\"at and
European Theoretical Spectroscopy Facility (ETSF), Max-Wien-Platz 1, 07743 Jena,  Germany }
\date{\today}
\begin{abstract}
Excitonic effects in optical spectra and electron-hole pair excitations are described by solutions of the Bethe-Salpeter equation (BSE) 
that accounts for the Coulomb interaction of excited electron-hole pairs. 
Although for the computation of excitonic optical spectra in an extended frequency range efficient methods are available, 
the determination and analysis of individual exciton states still requires the diagonalization 
of the electron-hole Hamiltonian $\hat{H}$. We present a numerically efficient approach for the calculation of 
exciton states with quadratically scaling complexity, which significantly diminishes the computational costs compared to 
the commonly used cubically scaling direct-diagonalization schemes. The accuracy and performance of this approach is demonstrated 
by solving the BSE numerically for the Wannier-Mott two-band model in {\bf k} space and the semiconductors MgO and InN. 
For the convergence with respect to the $\vk$-point sampling a general trend is identified, which can be used to extrapolate 
converged results for the binding energies of the lowest bound states.
\end{abstract}
\maketitle

\section{Introduction}
The Coulomb interaction of excited electrons and holes plays an important role for the optical properties of condensed matter 
\cite{Sham:1966:MPD,Hanke:1979:MPE,Strinati:1988:AGF}. Photon-induced two-particle electronic excitations are accompanied by the 
rearrangement of the remaining electrons in a solid, so that the individual particles are renormalized to quasiparticles (QPs).
Their description within the framework of many-body perturbation theory (MBPT)
has made substantial progress in the last three decades \cite{Onida:2002:EED}. 
The most common approach is Hedin's GW approximation \cite{Hedin:1965:F},
which describes the response of the remaining electrons by a dynamically screened Coulomb potential $W$. 
The self-energy operator $\Sigma$ of an excited particle is thereby given by $\Sigma=i\hbar GW$ with the single-particle Green's function $G$.
Its numerical implementation \cite{Hybertsen:louie:1985b,Godby:1987} usually yields single-particle excitation energies in good agreement 
with angle-resolved or inverse photoemission experiments \cite{Bechstedt:1992,Aryasetiawan:1998:GM,Aulbur:2000:SSP}.
However, the optically excited quasielectron-quasihole pairs show additional interactions.
They are described by the so called polarization function $P$, which obeys a Bethe-Salpeter equation (BSE) \cite{Sham:1966:MPD,Hanke:1979:MPE}. 
Apart from the bare electron-hole exchange, which can be identified with crystal local-field effects (LFEs) \cite{Hanke:1979:MPE}, 
its kernel is given by the derivative $-(i\hbar)^{-1}\delta\Sigma/\delta G$ which is usually replaced by the leading term $-W$. 
It clearly represents the screened Coulomb attraction between quasielectrons and quasiholes \cite{Strinati:1988:AGF}.

Optical spectra of real materials can be calculated from first principles by solving the
eigenvalue problem (EVP) for an effective two-particle Hamiltonian $\hat{H}$ corresponding to a reformulated BSE.
The eigensystem of $\hat{H}$ then can be used to obtain a spectral representation of $P$. 
Usually, one of the following numerical approaches is applied to calculate $P$ or the directly 
related frequency-dependent dielectric function $\epsilon(\omega)$: 
(i) The explicit diagonalization of $\hat{H}$  (solving the EVP directly \cite{Albrecht:1998:ICE,Rohlfing:1998:EHE}), 
(ii) the iterative Haydock method \cite{Benedict:1998:OAI}, or (iii) a time-evolution algorithm that is 
based on an initial-value formulation of the Fourier-transformed dielectric function \cite{Schmidt:2003:EMS}.
Meanwhile, the applications extend also to rather complex materials and structures such as semiconductor surfaces 
\cite{Rohlfing:1999:EOS,Hahn:2002:BEE}, nanocrystals and molecules \cite{Rohlfing:1998:EEO,Hahn:2005:MEE}, 
or even ice and water \cite{Hahn:2005:OAW,Garbuio:2006}. The most important changes in the optical spectra 
with respect to the independent-particle limit concern a general redshift
of the transition energies and a redistribution of oscillator strength towards lower energies \cite{Schmidt:2003:EMS}.

For some systems also bound states of the electron-hole pairs -- so called excitons \cite{Kittel:1996} -- form and can be 
observed in optical spectra below the QP absorption edge.
Thereby, two fundamentally different types of excitons -- the Frenkel and the Wannier-Mott type -- are distinguished.
Excitons of the Frenkel type emerge in systems where the gap is confined by rather localized electronic states,
as it is found for surfaces of covalently bonded semiconductors \cite{Rohlfing:1999:EOS}, insulators with strong ionic bonds
 \cite{Rohlfing:1998:EHE,Benedict:1998:OAI,Benedict:1999:ICI,Bechstedt:2005}, or crystals with hydrogen-bridge bonds \cite{Hahn:2005}. 
For materials with a pronounced dispersion of the first conduction band excitons of the Wannier-Mott type 
\cite{Kittel:1996,Yu:1999:FS,Elliot:1957,Shinada:1966} can form below the gap. 
They give rise to a hydrogen-like spectrum of pair eigenvalues and, in contrast to the Frenkel type, 
an electron-hole distance much larger than the lattice constant. 
Typically, their binding energies are much smaller than those of Frenkel-like excitons. 
Nevertheless, the lowest excitonic states can be probed by a variety of spectroscopic techniques. 

However, first-principles calculations, as outlined above, remain a challenging task especially for Wannier-Mott-like excitons with 
binding energies below 0.1~eV. 
The most important limitation is due to the high $\vk$-point densities required for sampling the Brillouin zone (BZ), 
in order to describe the localization of the excitonic wave function in the Fourier space sufficiently.
The number of $\vk$ points directly relates to the rank $N$ of the BSE Hamiltonian, causing extreme demands in 
terms of storage and CPU time for the calculation of the spectra. The latter one rapidly becomes the limiting 
factor, since, of the previously discussed approaches, only the direct diagonalization of $\hat{H}$ -- 
involving computational costs scaling like $\mathcal{O}(N^3)$ -- can be employed in the computation of bound pair excitations. 
Attempts have been made to solve the problem by introducing an interpolation scheme for the electron-hole interaction 
\cite{Rohlfing:2000:EHE} or by $\vk$-point samplings restricted to only a small part of the BZ 
\cite{Laskowski:2005,Laskowski:2006}.
Still, the question how the complete electronic structure influences the bound pair excitations and their the oscillator strengths remains open. 
Also the effects of electron-hole exchange on the excitons \cite{Denisov:1973,Fu:1999} 
have been studied little so far by a combination of {\em ab initio} electronic structure calculations and the 
BSE treatment of the two-particle excitations. Another important point concerns the prediction of excitonic effects, 
e.g. binding energies and oscillator strengths, for semiconductors such as InN, whose sample quality does currently 
not allow their measurement. Also the influence of polytypism of a material remains an open question.

In order to reduce the computational demands necessary for systematic studies addressing the aforementioned questions,
we combine the use of special $\vk$-point sets and an iterative matrix-diagonalization scheme, 
obtaining a computationally very efficient approach for the calculation of excitonic states below the absorption edge. 
The methodology of this approach is described in detail in Sec.~II. 
The computational details of the underlying {\em ab initio} calculations are summarized in Sec.~III, 
before in Sec.~IV.A the approach is applied to and tested by means of the Wannier-Mott model exciton.
The consequences of ''true''  {\em ab initio} band structures are studied for the semiconductors MgO and InN in Sec.~IV.B and C.
Finally, a summary and conclusions are given in Sec.~V.

\section{Methodology}
\subsection{Bethe-Salpeter equation and pair Hamiltonian}
The inclusion of excitonic effects in optical spectra requires to go beyond the independent-(quasi)particle approximation (IQA) 
for the polarization function $P$ by taking into account the electron-hole attraction and exchange. 
For this purpose a BSE for the irreducible polarization can be derived from MBPT \cite{Sham:1966:MPD,Onida:2002:EED},  
\begin{equation}
  P=P_0+P_0(2\bar{v}-W)P \label{equ:BSE},
\end{equation}
with the IQA polarization function $P_0$, the statically screened Coulomb potential $W$, and the bare 
Coulomb potential $\bar{v}$ without its long-range Fourier component $\vec{G}=0$. 
For a detailed derivation and a generalization for collinear spin polarization we like 
to refer to \cite{Roedl:2008}. 

A commonly adopted basis for representing the BSE is given by the Bloch states $\varphi_{n\vk}(\vr)$ of the crystal problem, 
characterized by the band index $n$ and a wave vector of the first BZ $\vk$. Neglecting QP effects for the wave functions,
they are usually approximated by the wave functions obtained in the framework of density functional theory (DFT)
from the solution of the Kohn-Sham equation,
\begin{equation}
  \left[-\frac{\hbar^2}{2m}\Delta+V(\vr)+V_H(\vr)+V_{XC}(\vr)\right]\varphi_{n\vk}(\vr)=\varepsilon_{n}(\vk)\varphi_{n\vk}(\vr),
\end{equation}
with the Kohn-Sham eigenvalues $\varepsilon_{n}(\vk)$ and the potentials of electron-ion ($V$), Hartree ($V_H$), and exchange and correlation 
(XC, $V_{XC}$) interaction. The latter potential is usually given by the local-density (LDA) or generalized-gradient approximation (GGA).
However, for materials where these approximations fail, such as InN, other potentials derived, e.g. from 
the LDA+U scheme  \cite{Anisimov:1991} or even spatially non-local potentials in the framework of generalized 
Kohn-Sham schemes \cite{Seidl:1996} might be used. Especially the latter constitute a good starting point for the pertubative 
calculation of single-particle QP eigenvalues $E^{QP}_{n}(\vk)$ \cite{Fuchs:2007:PRB}.

With restriction to materials with completely occupied and unoccupied bands (semiconductors and insulators), 
the BSE \eqref{equ:BSE} can be rewritten as EVP for an effective two-particle Hamiltonian $\hat{H}$ \cite{Albrecht:1998:ICE},
\begin{equation}
  \frac{\Omega}{(2\pi)^3}\sum_{c'v'}\int_{\Omega_{BZ}}d\vk'\hat{H}^{c}_{c'}{}^{v}_{v'}(\vk,\vk')A^{\Lambda}_{c'v'}(\vk')=
  E^{\Lambda}A^{\Lambda}_{cv}(\vk), 
  \label{equ:BSE_evp}
\end{equation}
with the pair eigenvalues $E^{\Lambda}$ and eigenvectors $A^{\Lambda}_{cv}(\vk)$,
$\Omega$ the crystal volume, and $\Omega_{BZ}$ the volume of the BZ.
The two-particle Hamiltonian can be divided into a diagonal and a non-diagonal part,
\begin{align}\label{equ:BSE_H}
\hat{H}^c_{c'}{}^v_{v'}(\vk,\vk')=H^D_{cv}(\vk)\delta_{vv'}\delta_{cc'}\delta_{\vk\vk'}+H^c_{c'}{}^v_{v'}(\vk,\vk'),
\end{align}
with
\begin{subequations}
\begin{align}
  H^D_{cv}(\vk)&=E^{QP}_{c}(\vk)-E^{QP}_{v}(\vk),\\
  H^c_{c'}{}^v_{v'}(\vk,\vk')&=-W^c_{c'}{}^v_{v'}{}^{\vk}_{\vk'}+2\bar{v}^c_{c'}{}^v_{v'}{}^{\vk}_{\vk'},
\end{align}
\end{subequations}
where $\delta_{\vk\vk'}$ is used as short notation for $(2\pi)^3\Omega^{-1}\delta(\vk-\vk')$.
The first, diagonal, contribution to \eqref{equ:BSE_H} is constructed from the difference of the QP eigenvalues of conduction and valence bands.
It describes the excitation of non-interacting quasiparticles.
The second part, given by the matrix elements of the statically screened Coulomb potential $W$, 
accounts for the attraction of electrons and holes,
\begin{equation}
\begin{split}
 W^{c}_{c'}{}^{v}_{v'}{}^{\vk}_{\vk'}&=
  \!\!\iint\! d\vr d\vr'\, \varphi^*_{c\vk}(\vr)\varphi_{c'\vk'}(\vr)W(\vr,\vr')\varphi_{v\vk}(\vr')\varphi^*_{v'\vk'}(\vr')\\
  &=\frac{1}{\Omega}\sum_{\vG\vG'}\frac{4\pi e^2\epsilon^{-1}_{\vG\vG'}(\vk-\vk')}{|\vk-\vk'+\vG||\vk-\vk'+\vG'|}
             B^{c}_{c'}{}^{\vk}_{\vk'}(\vG)B^{v}_{v'}{}^{\vk*}_{\vk'}(\vG'),
\end{split}
\label{equ:BSE_W}
\end{equation}
with the symmetrized inverse dielectric function $\epsilon^{-1}_{\vG\vG'}$ and the Bloch integrals
\begin{equation}
 B^{n}_{n'}{}^{\vk}_{\vk'}(\vG)=\frac{1}{\Omega_0}\int_{\Omega_0} d\vr\, u^*_{n\vk}(\vr) e^{i\vG\vr} u_{n'\vk'}(\vr).
\end{equation}
Thereby, $\vG$ denotes vectors of the reciprocal lattice, $\Omega_0$ the unit cell volume, and $u_{n\vk}(\vr)$ the cell-periodic part of the Bloch waves 
$\varphi_{n\vk}(\vr)=(\Omega)^{-1/2}e^{i\vk\vr}u_{n\vk}(\vr)$. 
The third part contains matrix elements of the bare Coulomb potential and models LFEs by means of 
an electron-hole exchange term \cite{Strinati:1988:AGF,Denisov:1973},
\begin{equation}
\begin{split}
\bar{v}^{c}_{c'}{}^{v}_{v'}{}^{\vk}_{\vk'}&=\!\!\iint\! d\vr d\vr'\, \varphi^*_{c\vk}(\vr)\varphi_{v\vk}(\vr)\bar{v}(\vr-\vr')
\varphi_{c'\vk'}(\vr')\varphi^*_{v'\vk'}(\vr')\\
&=\frac{1}{\Omega}\sum_{\vG\neq 0}\frac{4\pi e^2}{|\vG|^2}B^{c\vk}_{v\vk}(\vG)B^{c'\vk'*}_{v'\vk'}(\vG).
\end{split}
\label{equ:BSE_vbar}
\end{equation}

\subsection{Generalized eigenvalue problem}\label{sec:bse_gevp}
For any practical purpose the $\vk$-continuous formulation of the BSE \eqref{equ:BSE_evp} needs to be discretized  \cite{Rohlfing:2000:EHE} 
and the sum over the valence and conduction bands has to be truncated. 
In this work, the latter truncation is achieved by introducing a cutoff energy for the transition energies of non-interacting pairs. 
For practical reasons we define this BSE cutoff $E_{cut}$ with respect to the single-particle energies without QP corrections, 
corresponding to the Kohn-Sham eigenvalues, $\varepsilon_{c}(\vk)-\varepsilon_{v}(\vk)\leq E_{cut}$.
For the $\vk$ discretization the BZ is divided into subvolumes $V_{\vk}$ with $\Omega_{BZ}=\sum_{\vk}V_{\vk}$.
The $\vk$-discrete BSE Hamiltonian and eigenvectors are defined as averages over the subvolumes $V_{\vk}$,
\begin{equation}
  A^{\Lambda}_{cv\vk}:=\frac{1}{V_{\vk}}\int_{V_{\vk}} d\vq\, A^{\Lambda}_{cv}(\vq),\quad 
          H^D_{cv\vk}:=\frac{1}{V_{\vk}}\int_{V_{\vk}} d\vq\, H^{D}_{cv}(\vq),
  \label{equ:BSE_ddef1}          
\end{equation}
\begin{align}
  H^{c}_{c'}{}^{v}_{v'}{}^{\vk}_{\vk'}:=\frac{1}{V_{\vk}V_{\vk'}} \int_{V_{\vk}} d\vq\int_{V_{\vk'}} d\vq'\, H^{c}_{c'}{}^{v}_{v'}(\vq,\vq').
  \label{equ:BSE_ddef2}       
\end{align}
With these definitions the discretization of \eqref{equ:BSE_evp} yields the generalized EVP
\begin{equation}
  \frac{\Omega}{(2\pi)^3}\sum_{c'v'\vk'}H^{c}_{c'}{}^{v}_{v'}{}^{\vk}_{\vk'}V_{\vk'}A^{\Lambda}_{c'v'\vk'}=
  (E^{\Lambda}-H^D_{cv\vk})A^{\Lambda}_{cv\vk},
  \label{equ:BSE_gevp}
\end{equation}
Fortunately, it is easy to recast \eqref{equ:BSE_gevp} into a conventional EVP for 
$\tilde{H}^{c}_{c'}{}^{v}_{v'}{}^{\vk}_{\vk'}=\alpha_{\vk}H^{c}_{c'}{}^{v}_{v'}{}^{\vk}_{\vk'}\alpha_{\vk'}$,
\begin{equation}
  \sum_{c'v'\vk'}\tilde{H}^{c}_{c'}{}^{v}_{v'}{}^{\vk}_{\vk'}\tilde{A}^{\Lambda}_{c'v'\vk'}=(E^{\Lambda}-H^D_{cv\vk})
  \tilde{A}^{\Lambda}_{cv\vk},
  \label{equ:BSE_devp}
\end{equation}
using the transformed eigenvectors $\tilde{A}^{\Lambda}_{cv\vk}=\alpha_{\vk}A^{\Lambda}_{cv\vk}$ 
with $\alpha_{\vk}=\sqrt{V_{\vk}\Omega/(2\pi)^3}$. 

Special care has to be taken for the singular $\vG=\vG'=0$ part (SC) of $W$ appearing in the $\vk=\vk'$ 
term of ${H}^{c}_{c'}{}^{v}_{v'}{}^{\vk}_{\vk'}$:
 \begin{equation}  
  {\rm SC}[H^{c}_{c'}{}^{v}_{v'}{}^{\vk}_{\vk'}]=-\frac{4\pi e^2\epsilon^{-1}_{00}(0)}{\Omega V_{\vk}V_{\vk}} 
  \int_{V_{\vk}} d\vq\int_{V_{\vk}} d\vq' \frac{\delta_{vv'}\delta_{cc'}}{|\vq-\vq'|^2}. 
  \label{equ:BSE_SC}
\end{equation}
It can be evaluated analytically when the six-dimensional integration is approximatively replaced by the three-dimensional integral  
$\int_{S_{\vk}}d \vq/|\vq|^2$ over spheres $S_{\vk}$ of the volume represented by each $\vk$ point.
The singularity correction decreases with increasing number of $\vk$ points, converging to zero in the limit of vanishing $\vk-\vk'$ distance.
In the case of regular $\vk$-point meshes the singularity correction leads to a rigid shift of the 
eigenvalues to lower energies \cite{Puschnig:2002}.
For inhomogeneously distributed $\vk$-point sets, however, its value is not constant over all $\vk$ points.
Therefore, we do not adopt the spherical approximation in order to avoid spurious effects on the absolute values 
and relative positions of the eigenvalues. Instead, we carry out the six-dimensional integral \eqref{equ:BSE_SC} numerically, 
taking into account the actual size and shape of $V_{\vk}$. Since the numerical convergence of these 
integrals is utterly slow but perfectly linear when plotted over the sampling-point distance along one direction, 
we extrapolate the value of the singularity correction from the results of two calculations using 
$10^6$ and $20^6$ sampling points inside $V_{\vk}$. Still, the integration of \eqref{equ:BSE_SC} is rather time-consuming and therefore 
only sensible in case of a limited number of different sizes and shapes of $V_{\vk}$, 
e.g. in the case of regular or hybrid $\vk$ meshes as introduced in Sec.~\ref{sec:method_hybrid}.

\subsection{Optical spectra}
The diagonal components of the frequency-dependent macroscopic dielectric tensor $\epsilon^{jj}_{mac}(\omega)$ 
can be obtained using the eigenvectors $\tilde{A}^{\Lambda}_{cv\vk}$ and energies $E^{\Lambda}$ from the solution of \eqref{equ:BSE_devp},
\begin{equation}
\begin{split}
  \epsilon^{jj}_{mac}(\omega)=1+\frac{4\pi}{\Omega}\frac{e^2\hbar^2}{m_0}
  \sum_{\Lambda} \frac{f^{\Lambda}_{j}}{E^{\Lambda}} \sum_{\beta=\pm 1}\frac{1}{E^{\Lambda}-\beta\hbar(\omega+i\gamma)}, 
\end{split}\label{equ:bse_spectra}
\end{equation}
where the oscillator strengths $f^{\Lambda}_{j}$ are given by
\begin{align}
f^{\Lambda}_{j}=\frac{2}{m_0}\left|\sum_{cv\vk}\frac{\bra{\varphi_{c\vk}}p_{j}\ket{\varphi_{v\vk}}}
                                                {\varepsilon_{c}(\vk)-\varepsilon_{v}(\vk)}
                               \alpha_{\vk}\tilde{A}^{\Lambda *}_{cv\vk}
                        \right|^2 E^{\Lambda} 
\label{equ:osci_strength}
\end{align}
and $\bra{\varphi_{c\vk}}p_{j}\ket{\varphi_{v\vk}}$ denote the matrix elements of the momentum operator in the Cartesian direction $j$.

For the calculation of the dielectric function (DF), Schmidt {\it et al.} \cite{Schmidt:2003:EMS} 
devised an efficient approach, scaling like $\mathcal{O}(N^2)$ with the dimension $N$ of the excitonic Hamiltonian.
This approach avoids the direct diagonalization of $\hat{H}$ by reformulating \eqref{equ:bse_spectra} as an initial-value problem and solving that.
It, however, comes at the price of losing the explicit knowledge of the exciton excitation energies $E^\Lambda$ and 
wave-function coefficients $A^{\Lambda}_{cv\vk}$. Further, a certain broadening $\gamma$ 
of the optical transitions is mandatory for the numerical convergence
of the algorithm. For large and complex systems, where $N$ can reach up to $10^5$ or more, the computational costs for a full 
diagonalization of $\hat{H}$ become prohibitive due to their $\mathcal{O}(N^3)$ scaling. This renders the time evolution 
the only feasible approach for the calculation of the DF. Another method, also aiming at the DF only, was proposed by Benedict {\it et al.} 
\cite{Benedict:1998:OAI}. It is, however, hard to compare in terms of computational efficiency due to its implementation in real space.

\subsection{An $\mathcal{O}(N^2)$ algorithm for  excitons}\label{sec:method_algo}
\begin{figure}
  \includegraphics[width=\columnwidth]{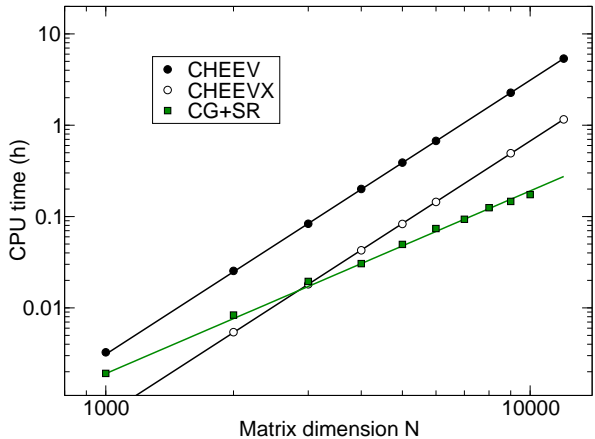}
  \caption{(Color online) CPU time on a single Opteron 275 core needed for the full diagonalization (CHEEV, black filled circles) or the 
           computation of the 15 lowest eigenvalues of randomly filled Hermitian matrices, 
           using the LAPACK library routine CHEEVX (black unfilled circles) or the CG+SR algorithm (green squares).
           The solid lines represent cubic (black) and quadratic fits (green) to the different data sets.}
  \label{fig:time.vs.dim}
\end{figure}
Often, not the DF itself but isolated exciton levels, especially bound states with pair excitation energies 
$E^{\Lambda}$ below the lowest QP gap, are of interest. Due to the limited spectral resolution and the possibility of 
vanishing oscillator strength, such excitations cannot be studied with the time-evolution method. 
The same holds for the analysis of excitonic wave functions, which requires the knowledge of the respective 
eigenvalues and eigenvectors.
In those cases Eq.~\eqref{equ:BSE_evp} needs to be solved directly, raising again the problem of 
excessive computational costs for complex systems.
For Wannier-Mott-like excitonic states near the absorption edge, usually a fairly small BSE cutoff can be used in order to reduce the 
dimension of $\hat{H}$ \cite{Schleife:2007:exciton}. On the other hand, a large number of $\vk$ points is required in the computations 
to ensure a sufficiently large crystal volume to accommodate the exciton, which can extend over several thousand unit cells according to 
exciton localization radii in the range of $10 - 100$ \AA. 

For excitons, numerical convergence is typically obtained for dimensions of $\hat{H}$ somewhere in the range of $N=10^4 - 10^5$. 
Direct matrix diagonalization schemes can be used at the lower end of the aforementioned range with computational times of a few CPU hours.
However, due to the rapid $\mathcal{O}(N^3)$ increase of computational costs, already problems of mid-range size are computationally too 
expensive for systematic studies. 

Given the combination of a large-rank Hamiltonian and the interest in only a few of the smallest eigenvalues 
and corresponding eigenvectors renders the problem strongly reminiscent of the Kohn-Sham problem in DFT.
Of course, this refers only to the dimensions of the problems, because in fact the BSE-EVP is less complicated, since no self-consistent 
update of the Hamiltonian is required.  
Using this observation,  we propose to employ iterative minimization techniques similar to those used in DFT for the study of distinct exciton levels. 
These are based on the minimization of the Ritz functional,
\begin{equation}
  \tilde{E}[\tpL]=\frac{\bra{\tpL}\hat{H}\ket{\tpL}}{\braket{\tpL}{\tpL}},
\end{equation}
with the trial vector $\ket{\tpL}$. Obviously, an unconstrained minimization yields
\begin{equation}
  \min_{\tpL}\tilde{E}[\tpL]=E^{\Lambda=1}=E[A^{\Lambda=1}],
\end{equation}
the lowest (first) eigenvalue and the corresponding eigenvector of $\hat{H}$. 
For any higher eigenvalue $\Lambda>1$ the minimization needs to be constrained to the orthogonal 
complement of ${\rm span}(A^1,\dots,A^{\Lambda-1})$.
For the actual calculation we follow the scheme devised by Kalkreuter and Simma \cite{Kalkreuter:1995}.
There, consecutive conjugate gradient (CG) steps are performed together with intermediate diagonalizations of 
$\hat{H}$ in the low dimensional subspace ${\rm span}(\tilde{\phi}^1,\dots,\tilde{\phi}^{n_s})$, so called subspace rotations (SR). 
The dimension $n_s$ of the subspace, thereby, roughly corresponds to the number of desired eigenvalues (typically $n_s\approx 20$). 
We have chosen a CG-based algorithm for the sake of simplicity and due to its robustness.
Other algorithms, however, can be used as well and might also improve performance. 

As already stressed above, in practice only a limited number of excitonic levels are of interest for the study of excitons.
Further, this number does usually  not depend on $N$, the dimension of $\hat{H}$. Therefore, under the assumption that 
the number of total CG steps is independent of $N$, the theoretical operation-count scaling of the 
CG+SR algorithm is found to be $\mathcal{O}(N^2)$.
The reason is that it involves only matrix-vector and vector-vector operations of dimension $N$ and 
diagonalizations of fixed-size $n_s\times n_s$ matrices.
In general, the assumption of a constant number of total CG steps, is of course, not true, since it is determined by the convergence of the 
algorithm. However, it is found that the convergence depends little on the matrix dimension, but more on its conditioning.  
Figure \ref{fig:time.vs.dim} confirms that the CG+SR method scales basically like $\mathcal{O}(N^2)$. The crossover point is found below
$N=1000$ when competing with full diagonalization, or at 3000 when comparing against the LAPACK \cite{LAPACK} CHEEVX routine for 
calculating selected eigenvalues and vectors of randomly filled Hermitian matrices. 

Moreover, the CG+SR algorithm can be easily parallelized. MPI parallelization can help to meet the high memory demands 
for solving the BSE by utilizing distributed memory. However, also very reasonable speedups are obtained by using multiple processors. 

\subsection{Hybrid $\vk$-point meshes}\label{sec:method_hybrid}
\begin{figure}
  \includegraphics[width=\columnwidth]{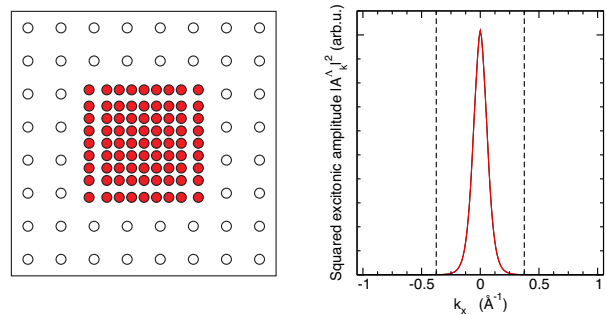}
  \caption{(Color online) The left panel illustrates the construction scheme for the hybrid $\vk$-point meshes for a two-dimensional BZ. 
           The $\vk$ points of the coarse mesh are indicated by unfilled circles, 
           those of the refined mesh by red dots. Further, the boundaries of the BZ and its inner part are indicated. 
           Following the notation introduced in the text, the mesh shown here would be referred to as: hybrid $8:3:21.33$.
           The right panel shows the localization of the wave function for an 1$s$ Wannier-Mott exciton (with the parameters as in Sec.~\ref{sec:res:wm}) 
           in reciprocal space.}
  \label{fig:hyb_mesh}
\end{figure}
Even though the algorithm suggested in the preceding section keeps the computational costs for the calculation of excitons tractable,
one drawback of the EVP remains -- its extreme memory demands. 
Using very dense $\vk$-point meshes the BSE Hamiltonian matrices \eqref{equ:BSE_H} easily exceed 50~GB storage. 
Since the storage requirements scale quadratically with the number of $\vk$ points, 
also the use of refined regular $\vk$-point meshes rapidly approaches the limits of today's supercomputers.

In order to reduce the memory demands, but also the computer time for setting up $\hat{H}$, 
one may benefit from noting that the wave functions of Wannier-Mott like excitons are well localized in 
$\vk$ space (cf.\ Fig.~\ref{fig:hyb_mesh}).
Therefore, it is reasonable to refine the $\vk$ meshes only in the center of the BZ \cite{Rohlfing:1998:EHE,Laskowski:2005}. 
Such a refinement, however, comes at the price of having to introduce varying weights 
(see Eqs.~(\ref{equ:BSE_ddef1})--(\ref{equ:BSE_gevp})), turning \eqref{equ:BSE_evp} into the generalized EVP \eqref{equ:BSE_devp}. 

The hybrid $\vk$-point meshes used in this work are constructed in the following manner (see also Fig.~\ref{fig:hyb_mesh}). 
Firstly, the whole BZ is sampled using a comparably coarse mesh of Monkhorst-Pack $\vk$ points \cite{Monkhorst:1976}. 
Secondly, an inner region, fenced in by $\vk$ points of the coarse mesh, is defined and the points inside 
this region including those on the borders are removed. 
In a third step, this inner region is filled with a fine-grained regular $\vk$ mesh, which 
must be constructed accordingly to include $\vk$ points on the borders of the inner region. 
Finally, the latter $\vk$ points are centered inside their associated $\vk$-space volume.
In some cases it is meaningful to further refine the $\vk$-point mesh in the inner region. 
This can be achieved by repeating steps two and three with respect to the inner mesh. 
Hereafter, meshes constructed accordingly by double repetition of the latter steps are referred to as double-hybrid ones. 

The following notation is used to characterize the afore-described hybrid meshes (see also Fig.~\ref{fig:hyb_mesh} and its caption).
The first set of numbers indicates the $\vk$ points in the outer (coarsely sampled) 
part of the BZ along the directions of the reciprocal basis.
The second set indicates the size of the central part of the BZ (with refined sampling) 
with respect to the $\vk$-point distance of the outer mesh,
and the third set of numbers provides a measure for the sampling density inside 
the central part of the BZ. For better comparison the latter set corresponds to the number 
of regular $\vk$ points in each direction necessary for sampling the whole BZ at the 
density achieved by the refined sampling. 
Double-hybrid meshes are characterized additionally by the size and sampling information of the second refinement region,
encoded in the same manner as for the single hybrid meshes described before.
To shorten the notation in case of isotropic sampling, only one number is given per set 
(see also the caption of Fig.~\ref{fig:hyb_mesh}).

One should note that the use of a hybrid mesh requires the knowledge of the 
dielectric screening $\epsilon^{-1}_{\vG\vG'}(\vq)$ at all vectors $\vq=\vk-\vk'$ 
for the  calculation of $W^{c}_{c'}{}^{v}_{v'}{}^{\vk}_{\vk'}$ in \eqref{equ:BSE_W}.
This also involves $\vq$ points that are not part of the hybrid $\vk$-point mesh,
which is no concern in case of a constant or model screening as used throughout this work.
However, if the screening is calculated in random-phase approximation (RPA), this constitutes an additional complication, that 
might be overcome by interpolating the screening for $\vq$ points not included in the hybrid mesh.

\section{Computational Details}
The results presented for MgO and InN in Sec.~\ref{sec:results}~B and C are based on the results of {\em ab initio}
DFT calculations, carried out using the Vienna {\em Ab initio} Simulation Package (VASP, \cite{Kresse:1996:EIT,Shishkin:2006}).
The projector-augmented wave (PAW) method \cite{Blochl:1994:PAW,Kresse:1998:UPP} is used to model the electron-ion interaction.
Plane waves up to a kinetic energy of 400~eV are included in the basis set of the electronic wave functions.
Quasiparticle band structures are calculated for InN and MgO as a gauge by means of the HSE03+G$_0$W$_0$ approach with  
computational details as described in detail in Refs. \onlinecite{Fuchs:2007:PRB,Shishkin:2006}.
Within the explicit calculations the single-particle eigenvalues and wave functions are computed using exchange and correlation 
potentials according to the semi-empirical LDA+U (InN) or the GGA (MgO) scheme 
together with a scissors operator to correct for the gap underestimation. 
The BSE Hamiltonian is set up according to \eqref{equ:BSE_H}, whereby, the inverse dielectric function entering the screened interaction 
term $W^{c}_{c'}{}^{v}_{v'}{}^{\vk}_{\vk'}$ is approximated by a $\vG$-diagonal model function \cite{Bechstedt:1992:EMC}.
The latter parametrically depends on the static electronic dielectric constant, which is taken from RPA calculations or experiment.

\section{Results}\label{sec:results}
In the following we present results obtained within the afore-introduced CG+SR approach for three applications. 
Firstly, we numerically study the two-band Wannier-Mott exciton in $\vk$ space to demonstrate the applicability of the 
CG+SR approach and to extract some general trends. 
Further, the formation of excitons is studied for the large-gap material MgO and the narrow-gap semiconductor InN.
Due to their band structures, the excitons occurring in both materials are expected to show Wannier-Mott-like character.

\subsection{The Wannier-Mott model}\label{sec:res:wm}
The probably most simplified model of a semiconductor band structure is the two-band model of two opposed parabolic bands,
separated by the fundamental gap $E_g$.
Within this model many fundamental properties can be calculated analytically, including the formation of a series of 
excitons below the absorption edge as found by Wannier \cite{Wannier:1937} and Mott \cite{Mott:1938}.
The analytic solution in three dimensions \cite{Elliot:1957} uses the fact that the BSE Hamiltonian \eqref{equ:BSE_H} for the two-band model,
\begin{equation}
  H^{c}_{c'}{}^{v}_{v'}{}^{\vk}_{\vk'}=\delta_{cc'}\delta_{vv'}\,\left[\left(E_g+\frac{\hbar^2|\vk|^2}{2\mu}\right)
                                       \delta_{\vk\vk'}-\frac{4\pi e^2}{\Omega\epsilon_0|\vk-\vk'|^2}\right],
\label{equ:WM_BSE_H}
\end{equation}
resembles that of a hydrogen-like atom with the known solution of a Rydberg series for the discrete part of the spectrum,
\begin{equation}
  E_{nlm}=E_g-R_{ex}/n^2,
\label{equ:WM_ev}
\end{equation}
where the quantum numbers $n,\ l$, and $m$ of the hydrogen atom now label the excitonic states.
Henceforth, $R_{ex}=R_{\infty} \mu/(m_0 \epsilon_0^2)$ denotes the excitonic Rydberg constant 
and $\epsilon_0$ determines the effective static and spatially constant screening of the Coulomb potential. 
The reduced mass $\mu=(m^{*-1}_c+m^{*-1}_v)^{-1}$ follows from the effective masses of the valence and conduction band.

For the numerical solution the effective masses are chosen to be $1.0\ m_0$ and $0.5\ m_0$, such that $\mu=1/3\ m_0$.
For the band gap a value of 3.0~eV is assumed. 
The calculations are performed in a simple cubic $\vk$-space volume extending over $2\pi/3$ \AA$^{-1}$, which naturally limits the maximum 
BSE cutoff to roughly 15.5~eV. Therefore, in the following only BSE-cutoff energies of 15~eV or below are considered. 
Using a screening constant of $\epsilon_0=4$, the excitonic Rydberg -- corresponding to the highest possible exciton binding energy -- 
amounts to $R_{ex}=283.45$~meV.
\begin{figure}
  \includegraphics[width=\columnwidth]{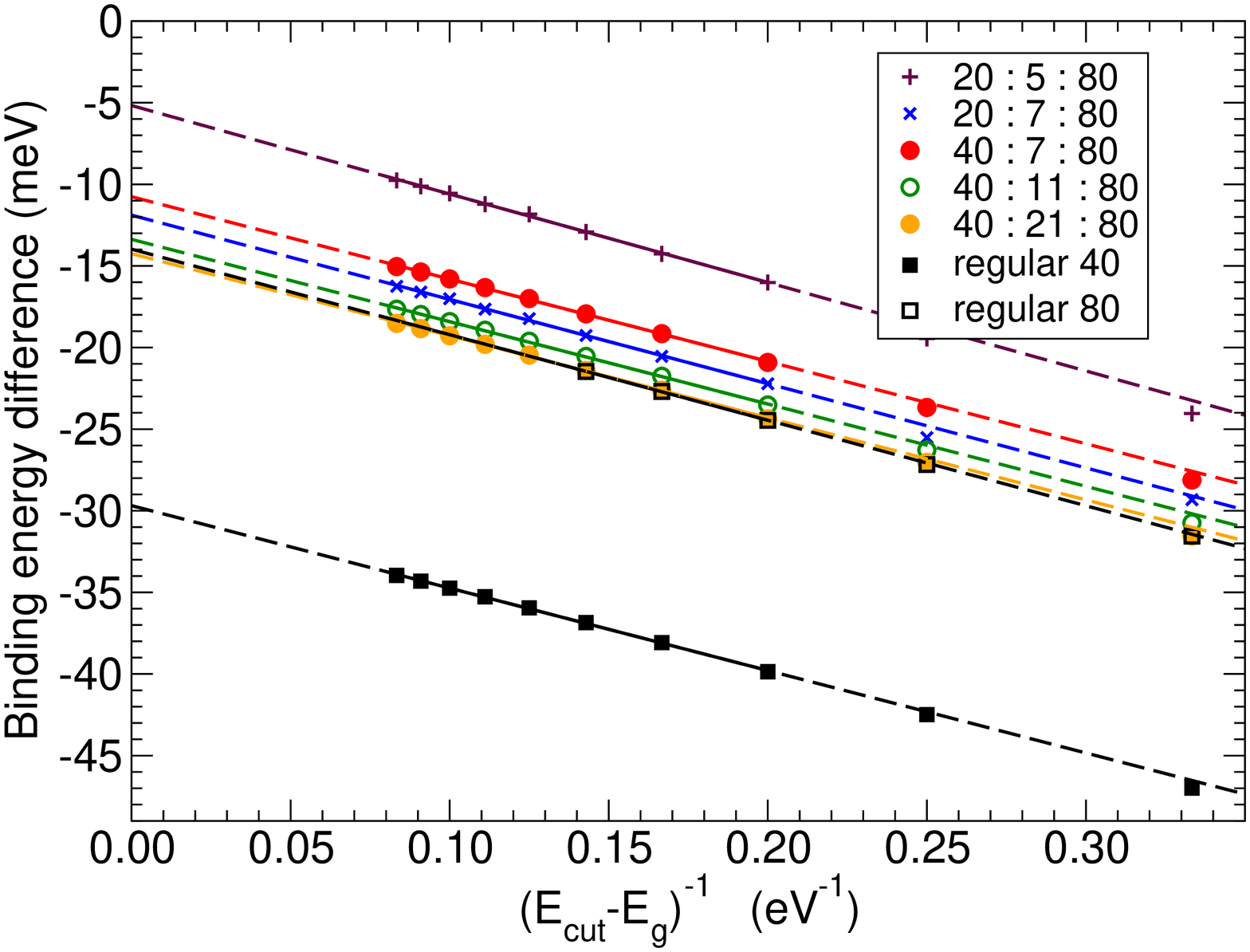}
  \caption{(Color online) Convergence of the binding energy of the 1$s$ exciton with the cutoff energy $E_{cut}$. 
           Shown are the results for two regular (squares) and five hybrid $\vk$-point meshes (circles and crosses).
           The hybrid meshes feature a common $\vk$ density in the inner part, equivalent to that of the regular  $80^3$ $\vk$-point mesh.
           The $y$ axis refers to the difference of the numerically calculated binding energy and the analytical result of 283.45~meV.
           Solid lines indicate the results of linear fits inside the region 0.08--0.2~eV$^{-1}$. 
           Dashed lines correspond to extrapolations based on these fits.}
  \label{fig:wm_Econ}
\end{figure}

In order to treat \eqref{equ:WM_BSE_H} numerically in $\vk$ space, it is necessary to introduce a $\vk$-point sampling. 
Using Monkhorst-Pack meshes of regular $\vk$ points, we first study the cutoff dependence of the binding energy 
of the 1$s$ ($n=1$, $l=0$) exciton. Figure \ref{fig:wm_Econ} shows the results for two different regular meshes consisting of 
$40\!\times\! 40\!\times\! 40$ and $80\!\times\! 80\!\times\! 80$ points.
According to Fig.~\ref{fig:wm_Econ} the 1$s$ binding energy increases with an increasing number of states included in $\hat{H}$ 
and a direct proportionality between the binding energy and $(E_{cut}-E_g)^{-1}$ exists.
While the latter relation cannot be expected to transfer to the situation of real semiconductors 
with a multitude of contributing bands,  it is  useful for estimating the effect of a limited BSE cutoff for the two-band model.
Based upon linear extrapolation of the data between $E_{cut}=8-15$~eV (cf.\ Fig.~\ref{fig:wm_Econ}), the 1$s$ binding energy at 15~eV cutoff 
is expected to underestimate the $E_{cut}\rightarrow\infty$ value by less than 5~meV, i.e. by less than 2\%.
The energy difference between the two $\vk$-point sets is much larger, demanding for a better $\vk$-point sampling.

Unfortunately, the $80\!\times\! 80\!\times\! 80$ $\vk$-point Hamiltonian matrix at a cutoff of 10~eV already requires 95~GB RAM in 
single precision storage and would require 480~GB at $E_{cut}=15$~eV, due to the relation $E_{cut}\propto N^{2/3}$,
specific for the two-band model. Therefore, 
the hybrid $\vk$-point meshes introduced in Sec.~\ref{sec:method_hybrid} are adopted in the following. 
The $\vk$-point density in the central part of the BZ corresponds to $80\!\times\! 80\!\times\! 80$ $\vk$ points 
in the full zone for all of them, thus allowing for a direct comparison to the results of the regular mesh with $80^3$ points.
According to Fig.~\ref{fig:wm_Econ}, it is possible to increase the distance of adjacent $\vk$ points by a factor of two in the outer zone 
without introducing any significant error, if the inner zone extends to half of the BZ along each direction.
Further, it should be noted that this $40 : 21 : 80$ mesh contains only about a fourth of the $\vk$ points included 
in the regular $80^3$ mesh and thereby reduces the memory demands by a factor of 16.
If the size of the densely sampled inner zone is reduced further, the binding energy increases. 
This is somehow unfortunate, because it counteracts the BSE-cutoff convergence. 
An analog trend is found, when the sampling in the outer region is reduced while the 
size and sampling density of the inner zone are kept fixed. 
Moreover, Fig.~\ref{fig:wm_Econ} shows that a trade-off between the minimum size of the inner zone and the sampling density of the outer zone
exists, if one  allows for a fixed error with respect to the corresponding regular mesh. 

\begin{figure}
  \includegraphics[width=\columnwidth]{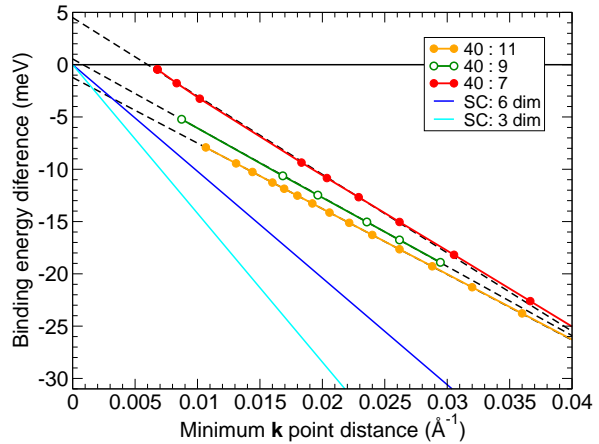}
  \caption{(Color online) Convergence of the $1s$ exciton binding energy with the $\vk$-point sampling in the inner zone 
           for three different sizes of the densely sampled inner zone and a fixed sampling density in the outer region of the BZ.
           The $y$ axis refers to the difference of the numerically calculated binding energy and the analytical result of 283.45~meV.
           The solid lines without symbols correspond to the singularity correction in the three-dimensional 
           spherical approximation (\cite{Puschnig:2002}, light blue) and the results from the 
           six-dimensional integration (cf.\ \eqref{equ:BSE_SC}, dark blue) 
           as used in the present calculations. }
  \label{fig:wm_k_con}
\end{figure}
Figure \ref{fig:wm_k_con} shows the results for the $1s$ binding energy using $\vk$ meshes with a fixed sampling density in the outer zone 
and a varying inner density and size of the inner zone.
Obviously, the binding energy is directly proportional to the distance of adjacent $\vk$ points, allowing for a linear extrapolation 
similar to the situation found for the convergence with the BSE cutoff. 
We would like to point out that the observed trend is not dominated by the singularity correction (cf.\ Eq.~\eqref{equ:BSE_SC}),
even though the latter obeys the same proportionality. This can be seen from Fig.~\ref{fig:wm_k_con}, which  
also includes plots of the singularity correction for the three-dimensional spherical approximation \cite{Puschnig:2002} 
and the six-dimensional integration \eqref{equ:BSE_SC} as utilized in the present calculations. Taking into account 
that the binding energy and not the eigenvalue itself is plotted, it can be seen from 
Fig.~\ref{fig:wm_k_con} that the singularity correction already partially anticipates the $\vk$-point convergence.

\begin{figure}[t]
  \includegraphics[width=\columnwidth]{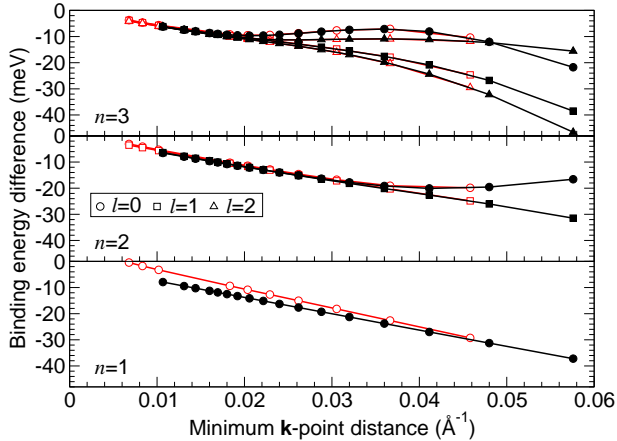}
  \caption{(Color online) Convergence of the first, second, and third shell ($n=1-3$) excitons with 
           the $\vk$-point sampling for two different sizes ($40:11$ - black lines with filled symbols and $40:7$ - 
           red lines with unfilled symbols) 
           of the densely sampled inner zone and a fixed sampling density in the outer region of the BZ.
           Different angular momentum quantum numbers $l$ are encoded by different symbols. Circles correspond to excitons with 
           $s$, squares to $p$, triangles to $d$ character. The binding energy is given with respect to the analytical 
           result of 283.45$/n^2$~meV with $n=1-3$.
          }
  \label{fig:wm_k_con2}
\end{figure}
Figure \ref{fig:wm_k_con2} shows the $\vk$-point convergence of the first, second and third shell excitons for two sizes of the inner zone.
In the realm of the two-band model all eigenvalues of a fixed principal quantum number $n$ are expected to be degenerate. 
Indeed, this degeneracy is found in the better-converged numerical results with a remaining error of about 0.1~meV or less. 
It is obvious from Fig.~\ref{fig:wm_k_con2}, that the rate of convergence might differ from one shell to another, 
but also within a shell of equal $n$, at least in the lower sampling regime. There also the linear behavior can be spoiled.
This is readily understood, taking into account that the respective wave functions differ in their real-space localization.
Therefore, since the exciton localization volume increases with the principal quantum number $n$, 
convergence with respect to the $\vk$-point sampling 
is achieved harder for higher excitonic states. The same holds for excitons of lower quantum number $l$ when
comparing them to others of the same shell $n$. 
Since these states become more localized in the reciprocal space, it is in principle possible to use 
hybrid meshes with an even smaller inner zone for computing the latter. 
This also becomes apparent from Fig.~\ref{fig:wm_k_con2} comparing the different $\vk$ meshes.

\begin{figure}[t]
  \includegraphics[width=0.8\columnwidth]{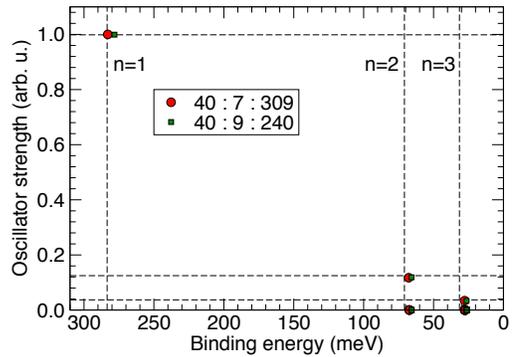}
  \caption{(Color online) Oscillator strengths of the $n=1-3$ excitons. Vertical and horizontal dashed lines indicate 
           the analytical results for the exciton binding energies and oscillator strengths. 
           Only excitons of angular momentum quantum number $l=0$ are found to be bright. All other excitons with $l>0$ are dipole-forbidden as
           indicated by vanishing oscillator strengths.}
  \label{fig:wm_osci}
\end{figure}
The oscillator strengths of the $n=1-3$ excitons are shown in Fig.~\ref{fig:wm_osci}.
From the analytic description only excitons with $s$ character are expected to have non-vanishing oscillator strengths,
which decay with increasing principal quantum number like $n^{-3}$.
Indeed the numerical results follow these expectations with only small deviations. 
Further, the $l>0$ excitons are found to have vanishing oscillator strengths.
This is in good agreement with the analytical result, predicting a direct proportionality between $f^{\Lambda}_{j}$ and 
the excitonic wave function at vanishing electron-hole distance in the case of $\vk$-independent dipole-matrix elements 
(cf.\ Eq.~\eqref{equ:osci_strength}). Since hydrogenic wave functions are non-vanishing at $r=0$ only for $l=0$, the $l>0$ 
excitons are dipole-forbidden.

\subsection{MgO}
\begin{figure}[t]
  \includegraphics[width=\columnwidth]{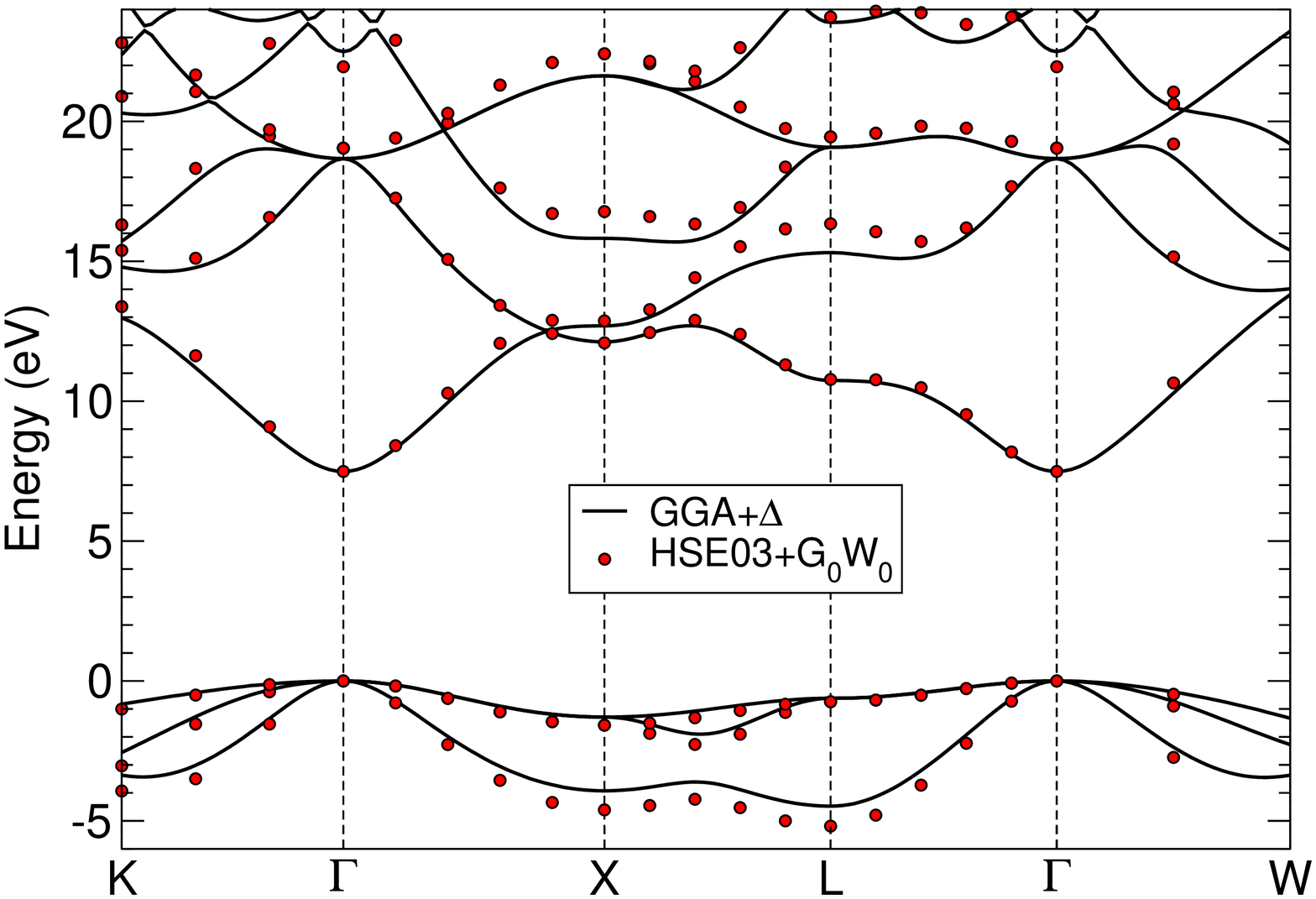}
  \caption{(Color online) MgO band structure in the GGA+$\Delta$
  (solid black lines) and HSE03+G$_0$W$_0$ (red circles)
  approximation. The scissors shift $\Delta$ corrects for the gap
  underestimation.}
  \label{fig:MgO_bands}
\end{figure}

Magnesium oxide (MgO) is a wide gap semiconductor
with a gap of approximately 7.8~eV~\cite{Roessler:1967}. Up to very
high pressures its equilibrium crystal structure is given by the
rocksalt ($rs$) phase~\cite{Schleife:2006}. In DFT-GGA the gap is
strongly underestimated with a value of only 4.5~eV. Using
first-order perturbation theory (G$_0$W$_0$) based on the DFT
eigenvalues, the gap cannot be fully corrected, resulting in a
GGA+G$_0$W$_0$ QP value of 6.8~eV. This deficiency can be mostly
overcome starting from the nonlocal HSE03 XC functional. Together with
perturbative QP corrections this approach results in a reliable
electronic structure with a slightly underestimated gap of 7.5~eV. 
Other, even more sophisticated QP calculations, taking into account self-consistency 
\cite{vanSchilfgaarde:2006,Shishkin:2007}
or even vertex corrections \cite{Shishkin:2007} in the self-energy, are found to 
overestimate the gap slightly by approximately the same amount.
All of these schemes, however, share the drawback of being computationally by far too expensive to act as 
foundation for a study of bound excitons.
Therefore, we use the GGA electronic structure as basis for the following studies,
simply correcting the gap by a scissors shift of 2.98~eV. Figure
\ref{fig:MgO_bands} demonstrates that this approximation compares
reasonably to the results of the more sophisticated HSE03+G$_0$W$_0$
approach. Especially the band dispersion is found to compare well between both approaches.

Due to the rocksalt symmetry of the crystal lattice, MgO has an $s$-like conduction-band minimum (CBM) of $\Gamma_{1,c}$-type and a threefold-degenerate
$p$-like valence-band maximum (VBM) of $\Gamma_{5,v}$-type. 
Two of the three uppermost valence bands show almost the same dispersion in the region around the
$\Gamma$ point, while the third one is more dispersive. In this situation
the formation of three degenerate, $s$-symmetric, excitons is expected from $\vk\cdot\vec{p}$ theory \cite{Baldereschi:1971}.
In the following we will label these excitons by A, B, and C.
Indeed, for well converged $\vk$-point samplings A, B, and C are found to be almost degenerate in our numerical calculations.
The remaining splitting between (A, B) and C amounts to 0.30~meV, corresponding to less than 0.1\% of the total binding energy,
and might be due to numerical errors. 
In full agreement with the initial expectations all three excitons are visible in any polarization direction. 
For the screening in our calculations we use an electronic dielectric constant of $\epsilon_{\infty}=3.0$ which is in between 
the experimental value of 2.94~\cite{Martienssen:2005} and the value of 3.16 obtained in RPA \cite{Schleife:2006}. 
The influence of $\epsilon_{\infty}$ will be discussed below.

\begin{figure}[t]
  \includegraphics[width=\columnwidth]{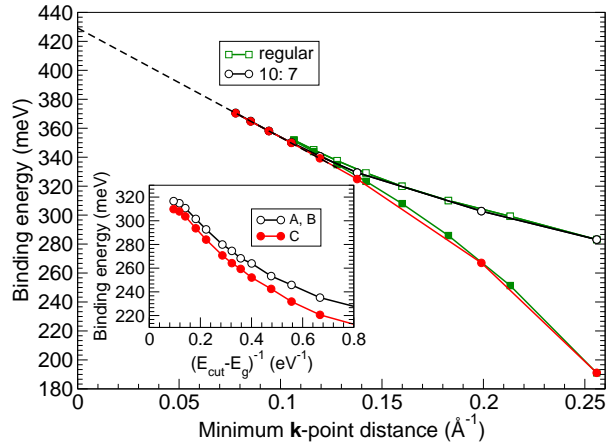}
  \caption{(Color online) Convergence of the MgO A, B (unfilled symbols), and C excitons (filled symbols)
           with respect to the $\vk$-point sampling and the BSE-cutoff energy (inset).
           The cutoff convergence was studied using a regular mesh of $16\!\times\! 16\!\times\! 16$ $\vk$ points,
           corresponding to a minimum $\vk$-point distance of 0.16~\AA$^{-1}$.
           Unfilled symbols refer to the binding energies of the A and B exciton, while filled symbols indicate those of the C exciton.
           The values of the hybrid mesh $10:7$ are extrapolated from the region of linear variation towards zero $\vk$-point spacing. 
          }
  \label{fig:MgO_convergence}
\end{figure}

In the inset of Fig.~\ref{fig:MgO_convergence} the convergence with respect
to the BSE-cutoff energy is shown for the lowest three excitons. Clearly, 
the variation is not strictly linear as in the case of the two band model. 
However, at BSE-cutoff energies higher than 13~eV, corresponding to $(E_{cut}-E_g)^{-1}=0.117$ eV$^{-1}$
the variation flattens out. Therefore, upon the data of Fig.~\ref{fig:MgO_convergence}, one can estimate that the exciton
binding energies computed at a BSE-cutoff energy of 13.0~eV, as used in the following, 
are accurate within about 10~meV, i.e., less than 2.4\% . 

The convergence with the number of $\vk$ points for MgO is studied in
Fig.~\ref{fig:MgO_convergence}. As found for the two-band model in the preceding section, we observe a linear variation with
respect to the distance of neighboring $\vk$ points. From this data
an average binding energy for the A, B, and C excitons of 429~meV is derived by linear extrapolation.
In comparison to experimental results of about 80~meV \cite{Roessler:1967} and 145~meV \cite{whited73}
the calculated binding energies turn out to be drastically overestimated. 
This may be attributed to the neglect of dynamical screening \cite{Shindo:1970,Zimmermann:1971}, 
which can influence the exciton binding energies significantly. 

Since the experimental binding energies are of the order of the longitudinal optical phonon energies $\hbar\omega_{LO}=89$~meV \cite{Martienssen:2005}, 
the lattice polarization can partially contribute to the screening of the electron-hole attraction. 
It is suggested \cite{whited69,Zimmermann:1971,Bechstedt:1980,Hahn:2005:MEE} that the static 
electronic dielectric constant $\epsilon_{\infty}$ has to be replaced by an effective dynamical one $\epsilon_0$, which is enlarged by parts
of the lattice polarizability $\approx(\epsilon_s-\epsilon_{\infty})$. 
For MgO the effect of lattice polarization is very pronounced, as indicated by the large difference between the 
static electronic dielectric constant $\epsilon_{\infty}=3.0$ and the static dielectric constant $\epsilon_s=9.8$ \cite{Martienssen:2005}.
This may lead to a strong reduction of the binding between electron and hole.
An estimate of the true effect would require an explicit treatment of the dynamical screening \cite{Shindo:1970,Zimmermann:1971}, which
is a computationally extremely demanding task. Therefore, we performed test calculations using an effective screening constant of $\epsilon_0=6.0$, 
as suggested from fits to the experimental exciton spectra \cite{whited69}, 
finding reduced exciton binding energies of in average 99.8~meV for the A, B, and C exciton.

It may be confusing to note that other theoretical studies \cite{Benedict:1998:OAI,Wang:2004} 
also predict much lower exciton binding energies closer to the experimental values, 
even though they use the static electronic dielectric constant or a RPA screened potential as well.
We attribute most of the difference to the comparably coarse $\vk$-point samplings of 256 \cite{Wang:2004} and 1000 \cite{Benedict:1998:OAI} 
$\vk$ points in the full BZ used in these studies to calculate the dielectric function in an extended energy range.
As demonstrated in Fig.~\ref{fig:MgO_convergence} the exciton binding energies of MgO show a rather strong variation with the $\vk$-point 
sampling density, leading to binding energies of 282~meV (A, B) and 191~meV (C) at a sampling density of 1000 $\vk$ points in the BZ. 

\begin{figure}[t]
  \includegraphics[width=\columnwidth]{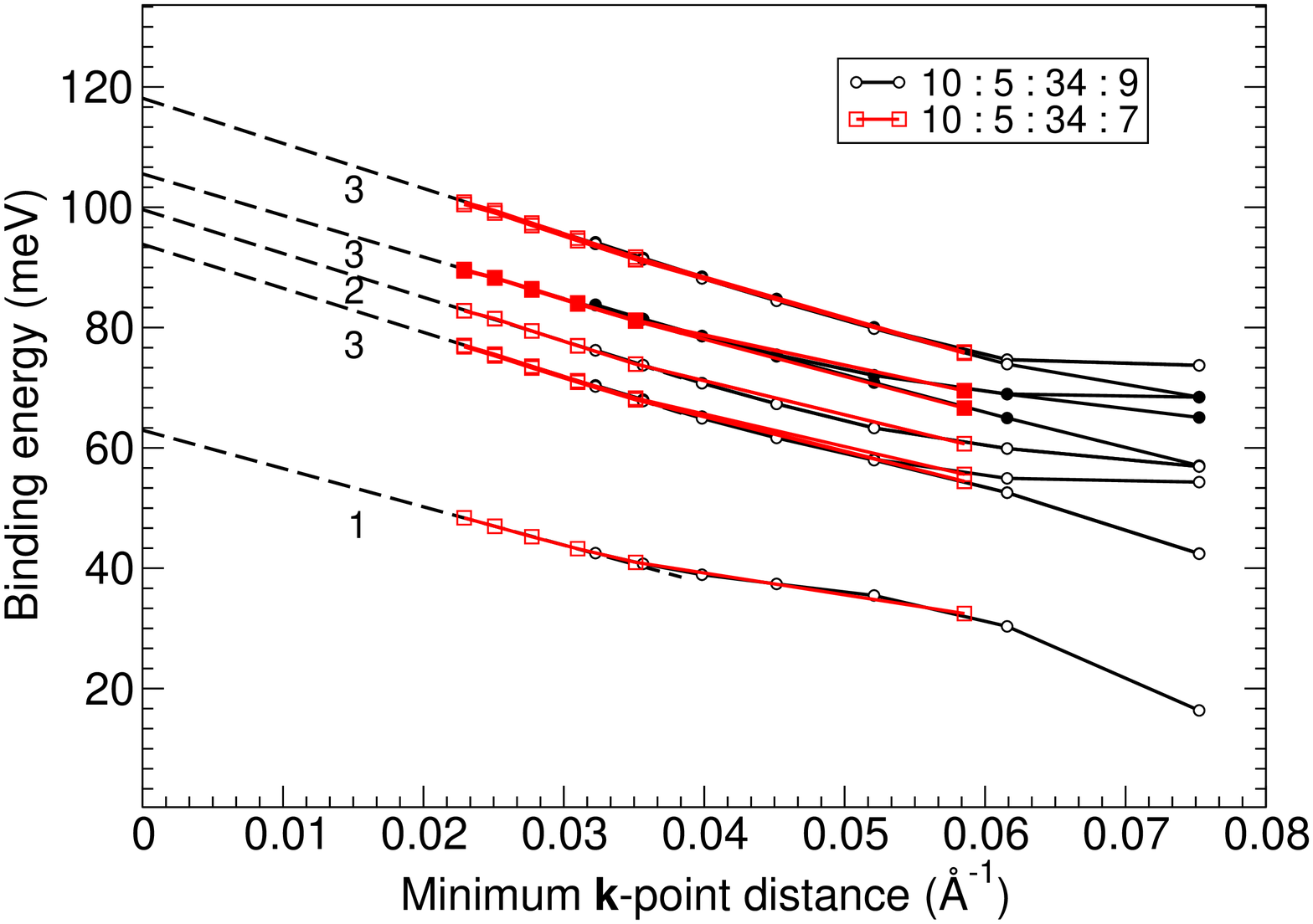}
  \caption{(Color online) Convergence of the second shell MgO excitons with respect to the $\vk$-point sampling.
           Filled symbols refer to bright excitations, while the excitons marked by unfilled symbols have vanishing oscillator strength. 
           The values of the double-hybrid mesh $10:5:34:7$ are extrapolated towards zero $\vk$-point spacing. Further, the degeneracies of the different eigenvalue complexes are annotated.
          }
  \label{fig:MgO_convergence2}
\end{figure}
An interesting effect of the real behavior of the electron-hole attraction can be observed for the higher pair excitation states.
There, the non-spherical deviations of the potential $W$ from the model $|\vk-\vk'|^{-2}$ dependence, are found to 
cause a splitting of the exciton levels 4--15 into five distinct levels. Three of the latter are threefold-degenerate,
and one of them is found to contain the only three excitons with non-vanishing oscillator strength (cf.\ Fig.~\ref{fig:MgO_convergence2}).
Figure \ref{fig:MgO_convergence2} shows the convergence of these states with respect to the distance of adjacent $\vk$ points, whereby
the linear variation at small $\vk$-point distances indicates a reasonable convergence. 

An open question is the absolute and relative influence of the LFEs or electron-hole exchange effects proportional to $\bar{v}$ 
in the pair Hamiltonian $\hat{H}$ \eqref{equ:BSE_H} on the binding of the excitons.
In order to study this effect, we setup the BSE Hamiltonian according to \eqref{equ:BSE_H}, once including both contributions 
($W$ and $\bar{v}$) and with $W$ or $\bar{v}$ separately. Two $\vk$-point samplings, 
given by the hybrid meshes $10:7:24$ and $10:7:20$, are used to account also for potentially different 
rates of convergence between the two contributions $W$ and $\bar{v}$.
As expected, excitonic bound states cannot form in absence of the attractive screened Coulomb interaction $W$ between electron and hole.
If, however, the electron-hole exchange $\bar{v}$ is suppressed, the binding energies of the lowest three excitons increase by about 58~meV.
In relation to the exciton binding energy the reduction due to the LFEs amounts to about 14\% or even more if dynamical 
screening is taken into account. The splitting of A, B,  and C is slightly affected. 
From Table \ref{tab:MgO} it can be inferred, that the convergence rates of $W$ and $\bar{v}$ indeed differ. 
Actually, the contribution of $\bar{v}$ is found to be well converged at the probed $\vk$-point sampling densities.
\begin{table}
\begin{ruledtabular}
\begin{tabular}{cccc}
\multirow{2}{*}{Hamiltonian}             &  \multicolumn{3}{c}{average A, B, C binding energy (meV)}  \\
                                         & extrapolated & $10:7:24$ & $10:7:20$ \\
\hline
{$\hat{H}=-W+2\bar{v}$}   & 429.2 & 370.5 & 358.1 \\
{$\hat{H}=-W $}           & 479.6 & 419.9  & 407.3  \\
\end{tabular}
\end{ruledtabular}
\caption{Effect of the electron-hole exchange on the average exciton binding energy of the  A, B,  and C excitons in MgO.}\label{tab:MgO}
\end{table}

\subsection{InN}\label{sec:results_inn}
Indium nitride (InN) is a narrow-gap semiconductor with a gap of approximately 0.6--0.7~eV \cite{Davydov:2002,Matsuoka:2002,Wu:2002},
crystallizing in  the wurtzite ({\it wz}) structure under ambient conditions.
Complications for the theoretical treatment of InN arise from the fact that DFT calculations 
including the In~$4d$ electrons as valence states 
fail to predict a finite gap, no matter whether a local (LDA) or semilocal (GGA) approximation is used for the  XC functional. 
Instead, the calculated band structures correspond to that of a zero-gap semiconductor with an, 
in comparison to the experiment, inverted ordering of the $\Gamma_{1,c}$ and $\Gamma_{5,v}, \Gamma_{1,v}$ levels \cite{Furthmuller:2005}. 
It was shown recently \cite{Rinke:2006,Fuchs:2007:PRB} that these deficiencies can be overcome, 
using either the method of optimized effective potentials \cite{Rinke:2006} or a hybrid XC functional like HSE03 \cite{Heyd:2003}.
Both methods, however, are computationally too expensive for a study of excitons.

\begin{figure}[t]
  \includegraphics[width=\columnwidth]{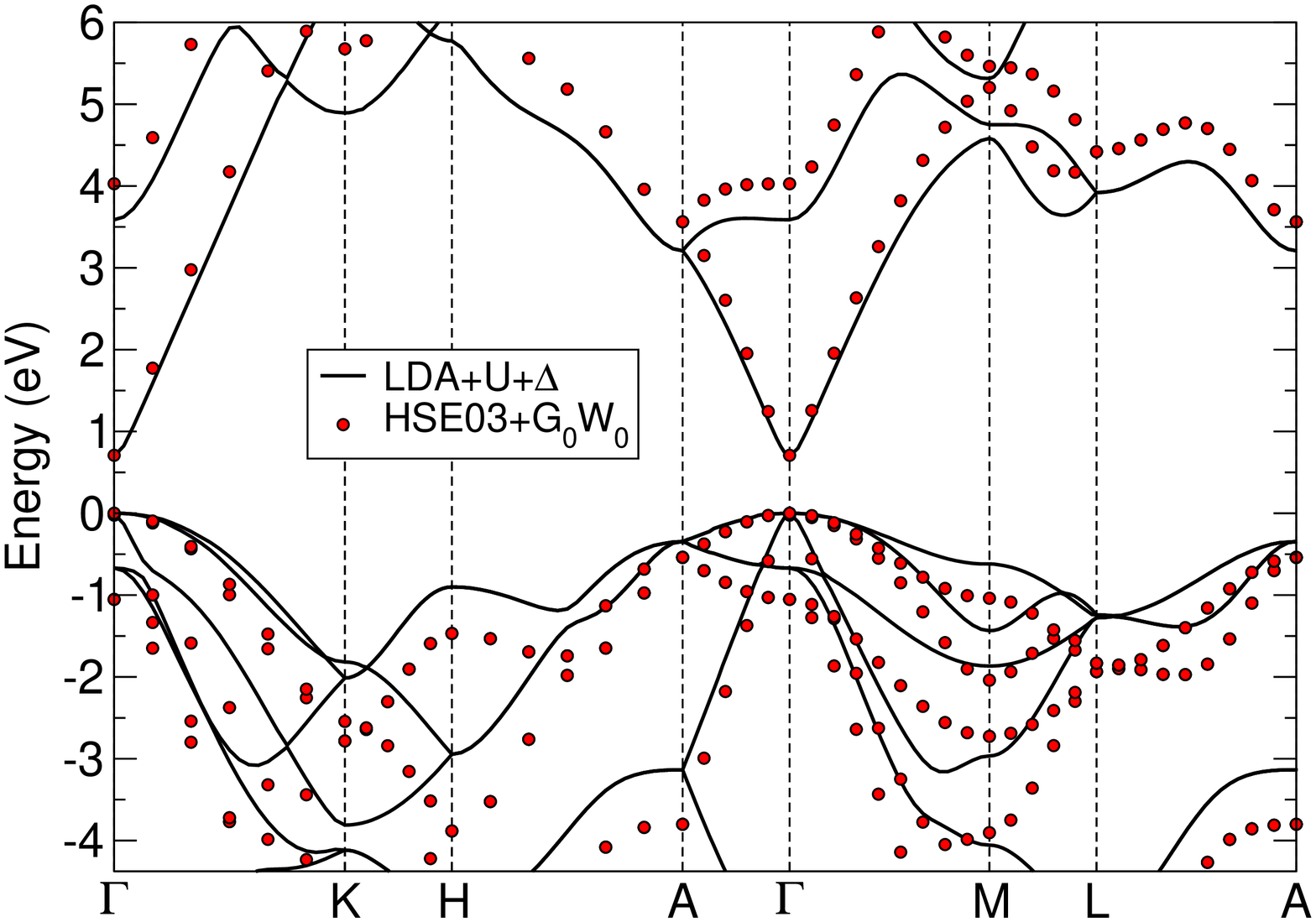}
  \caption{(Color online) InN band structure in the LDA+U+$\Delta$ (solid black lines) and HSE03+G$_0$W$_0$ (red circles) approximation.
           A U parameter of 3~eV is used for a good approximation of the HSE03+G$_0$W$_0$ band structure in the gap region. 
           The scissors shift $\Delta$ corrects for the remaining gap underestimation. }
  \label{fig:InN_bands}
\end{figure}
Therefore, we retreat to the much simpler LDA+U scheme \cite{Anisimov:1991,Anisimov:1993,Franchini:2005},
trying to obtain a reasonable approximation of the HSE03+G$_0$W$_0$ band structure in the gap region 
around the $\Gamma$ point by tuning the intra-atomic Coulomb repulsion of the In~$4d$ electrons by the U parameter.
The band structure resulting from such an LDA+U calculation is shown in Fig.~\ref{fig:InN_bands} 
in comparison to the QP band structure derived within the HSE03+G$_0$W$_0$ scheme. 
For clarity only the results fitting best, obtained for U=3~eV, are shown. 
Moreover, a scissors shift of 0.48~eV is used to open the gap additionally towards the 
value of 0.71~eV, as predicted by the HSE03+G$_0$W$_0$ calculations. 
Using the LDA+U approach, the effective masses at the $\Gamma$ point amount to approximately $2.2-2.5\ m_0$ 
for the uppermost valence bands ($v_A, v_B$) of $\Gamma_{5,v}$ type. The masses of the crystal-field split off ($v_C$, $\Gamma_{1,v}$) 
and the first conduction band are much smaller with values of approximately $0.03\ m_0$. 
The respective values obtained from the HSE03 calculation differ notably, amounting to $0.9\ m_0$ for the mass of the $v_A$ and $v_B$ bands and $0.046\ m_0$  
for the $v_C$ band and electron mass. We will try to assess the influence of the modified dispersion at the end of this section.
The crystal-field splitting amounts to $\Delta_{cf}=$11.9~meV in the present approximation of LDA+U (U=3~eV), 
underestimating slightly the experimental values of $19-24$~meV \cite{Goldhahn:2006} but also the HSE03+G$_0$W$_0$ value of 22.9~meV. We do not take into account the spin-orbit splitting of the valence bands, which amounts to 21.3~meV 
according to HSE03 calculations \cite{Fuchs:2008:unpub}.
For the static electronic dielectric constant $\epsilon_{\infty}$ entering $W$ a value of 7.9 is used. It corresponds 
to the average of the experimental values of the ordinary (7.83) \cite{Schley:2007} and extraordinary (8.03) \cite{Goldhahn:2004} 
polarization component determined from the analysis of the respective DF below the {\it wz}-InN gap of 0.68~eV.

Due to the wurtzite symmetry and the neglect of spin-orbit interaction, the VBM of InN is given by a twofold degenerate state.
Therefore, the two strongest bound excitons (A, B), corresponding to the Wannier-Mott 1$s$ excitons, are expected to be twofold degenerate as well.
A third 1$s$-like exciton (C) is expected to form between the first conduction and the crystal-field split-off $v_C$ band,
at an eigenvalue differing from that of A and B by roughly $\Delta_{cf}$.
Since optical transitions between $v_A, v_B$ and the CBM at $\Gamma$ are dipole-allowed only in ordinary polarization ($\vec{E}\perp \vec{c}$)
and $v_C$-CBM transitions only in extraordinary polarization ($\vec{E}\parallel \vec{c}$) \cite{Chuang:1996}, the corresponding excitons are expected to 
have non-vanishing oscillator strengths in the respective directions \cite{Thomas:1959}. 

\begin{figure}[t]
  \includegraphics[width=\columnwidth]{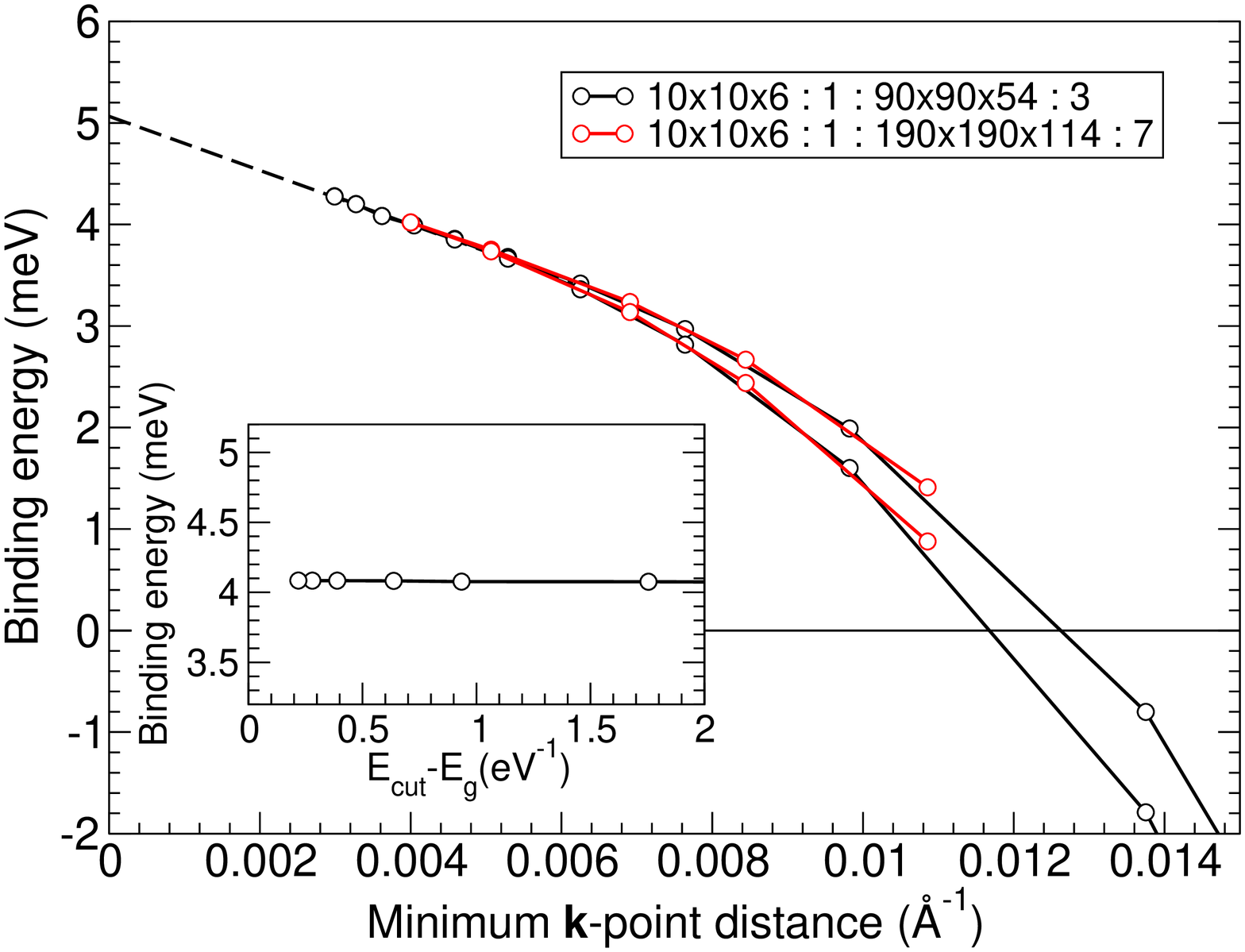}
  \caption{(Color online) Convergence of the InN A and B excitons with respect to the $\vk$-point sampling and the BSE-cutoff energy (inset).
           The cutoff convergence was studied using the double hybrid $\vk$-point mesh at a $\vk$-point distance of 0.0036 \AA$^{-1}$.
           The values of the double-hybrid meshes $10\!\times\! 10\!\times\! 6 : 1 : 90\!\times\! 90\!\times\! 54 : 1 $ are extrapolated towards zero $\vk$-point spacing. }
  \label{fig:InN_convergence}
\end{figure}

For a moment we limit ourselves to the discussion of the A and B excitons, which allow for a simple definition of their binding energies 
\cite{Laskowski:2005}. In agreement with the initial expectations  A and B are found to be degenerate. 
The inset of Fig.~\ref{fig:InN_convergence} shows their convergence with respect to the BSE-cutoff energy. 
The overall variation of the binding energy with respect to the BSE cutoff is extremely weak, due to the very small 
binding energies and the strong localization of the lowest conduction band around the $\Gamma$ point (cf.\ Fig~\ref{fig:InN_bands}). 
According to Fig.~\ref{fig:InN_convergence}, a BSE cutoff of 2~eV -- as used in the following -- should introduce an 
error of less than 0.1~meV.  

The convergence with the number of $\vk$ points is also demonstrated in Fig.~\ref{fig:InN_convergence}.
Due to the combination of a narrow gap, large screening, and a very small effective electron mass found for InN,
extremely dense $\vk$-point meshes are required to converge the exciton binding energies. 
Therefore, the double hybrid $\vk$-point meshes, as introduced in Sec.~\ref{sec:method_hybrid}, 
are used to study the formation of excitonic states in InN. 
At the high-sampling limit of the studied $\vk$-point meshes the linear variation with respect 
to the distance of adjacent $\vk$ points, as observed for the two-band model and MgO, is restored.
Extrapolating the binding energies of the lowest excitons towards zero $\vk$-point distance, 
reveals binding energies of about 5.0~meV for the A and B exciton. 
The absolute difference between the computed values and the value of $6.5$~meV, estimated using the two-band model together with 
the appropriate masses and screening constant, is small. 

\begin{figure}[t]
  \includegraphics[width=\columnwidth]{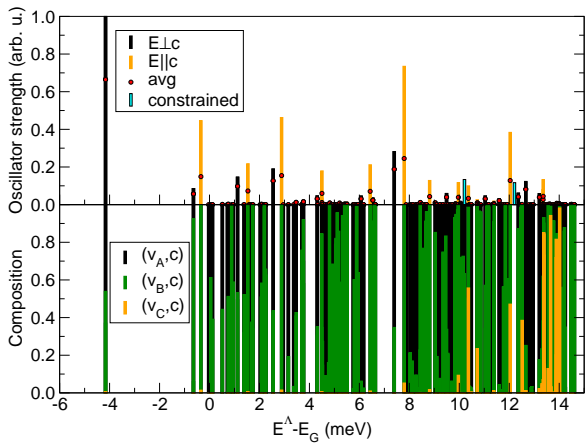}
  \caption{(Color online) Exciton spectrum of InN. The pair energies are given with respect to the lowest gap energy. 
           The oscillator strengths are given in the upper panel. Black solid bars indicate the average oscillator 
           strengths in ordinary polarization direction, while orange bars indicate the extraordinary polarization direction.
           The cyan bars indicate eigenstates of a pair Hamiltonian, which includes only contributions of the $v_C$ and the first conduction band.
           The red dots indicate the average oscillator strength. In the lower panel the contributions resulting from transitions between 
           $v_A, v_B, v_C$ and the first conduction band c are analyzed (see text).}
  \label{fig:InN_oscistrength}
\end{figure}
Figure \ref{fig:InN_oscistrength} shows the exciton spectrum of InN together with the respective oscillator strengths for ordinary 
and extraordinary light polarization. Further, the contributions $c^{\Lambda}_{vc}= \sum_{\vk} |\tilde{A}^{\Lambda}_{cv\vk}|^2$ 
of the inter-band transitions between the 3 uppermost valence bands $v_A$, $v_B$, $v_C$  and the first conduction band $c$ have been analyzed. 
They are indicated by the relative lengths of the differently colored bars in the lower panel of Fig.~\ref{fig:InN_oscistrength}.
Obviously, as the respective contributions sum up to one, only the three single-particle 
pair states $(v_A,c)$, $(v_B,c)$, and $(v_C,c)$ are found to contribute to the excitons in the energy interval shown in Fig.~\ref{fig:InN_oscistrength}. 
Now, also the C exciton can be identified. To this end, we perform a constrained calculation with a pair Hamiltonian 
including only $(v_C,c)$ transitions. The lowest eigenvalue of this Hamiltonian is found 2.3~meV below the $v_C$-CBM gap 
(cf.\ Fig.~\ref{fig:InN_oscistrength}). 
Indeed, the unconstrained pair Hamiltonian gives rise to an eigenvalue at an only marginally higher energy, 2.1~meV below the $v_C$-CBM gap.
The respective exciton, however, is found to be a nearly half-half mixture of $(v_B,c)$ and $(v_C,c)$ contributions. 
Further, it is clear from Fig.~\ref{fig:InN_oscistrength} that the latter exciton is neither the strongest bound state with contributions from $(v_C,c)$
transitions, nor it is the first exciton visible in extra-ordinary polarization, nor it has the strongest oscillator strength in the latter polarization
direction. Especially the latter two points are notable, since they contradict the initial expectations and may affect the experimental assignment
of excitons. The reason for the observed behavior is found in the $\vk$-point dependence of the single-particle optical oscillator strengths $\propto |\bra{\varphi_{c\vk}}p_{j}\ket{\varphi_{v\vk}}|^2$. 
While the optical transitions $(v_A,c)$ and $(v_B,c)$ at $\Gamma$ are dipole-forbidden in extraordinary polarization, they are allowed and 
of comparable strength with the $(v_C,c)$ transitions at off-$\Gamma$ $\vk$ points.

Finally, we address the influence of the LDA+U approximation on the present results. 
The electron and $v_C$ effective masses at $\Gamma$ increase with increasing U,
but underestimate the values calculated upon the HSE03 functional. 
Larger values than U=3~eV, however, yield a vanishing or even inverted crystal-field splitting 
-- contradicting the HSE03 results and experimental findings. 
Due to the accompanying crossing and mixing of the $v_A$, $v_B$, and $v_C$ bands, values higher than U=3~eV are not meaningful starting points.
The same holds for lower values of U, which predict much too small gaps and electron masses.
The effect of the underestimated electron mass can be estimated within the two-band model, 
where the binding energy is found to depend linearly on the reduced effective mass. 
Therefore, the computed binding energies of the A and B excitons may underestimate the actual values by about 30\%.

Apart from the influence of the U parameter the calculated exciton binding energies are influenced by the screening.
Especially dynamical screening \cite{Shindo:1970,Zimmermann:1971} may reduce the exciton binding. 
The characteristic optical phonon energies ($\hbar\omega_{LO}=86$~meV \cite{Martienssen:2005}) 
are much larger than $R_{ex}$, so that the lattice polarization can fully contribute to the screening
of the electron-hole attraction. For InN the static dielectric constant amounts to $\epsilon_s=13$ \cite{Davydov:1999}.
Using $\epsilon_s$ for the screening instead of $\epsilon_{\infty}$ would significantly reduce the exciton binding energies to values below 2~meV.
Consequently, Wannier-Mott excitons cannot be observed in the currently available InN samples, due to  
their high density of free carriers which is above the Mott-transition energy \cite{Davydov:2002,Matsuoka:2002,Wu:2002}.

\section{Summary and Conclusions}
We have introduced a new and highly efficient numerical approach for the solution of the homogeneous 
BSE for bound electron-hole pair excitations in non-metals.
The use of an iterative diagonalization scheme is demonstrated to diminish the computational costs for the calculation of 
bound electron-hole pair states significantly, allowing for the systematic study of excitonic Hamiltonians with ranks 
up to several hundred thousands. This is due to its, in comparison  to the $\mathcal{O}(N^3)$ scaling of direct diagonalization methods, 
favorable $\mathcal{O}(N^2)$ scaling.
Further, a hybrid $\vk$-space sampling scheme is presented which allows a refined sampling in selected parts 
of the BZ and simultaneously avoids an artificial localization of the exciton wave functions. 
The corresponding discretization of the pair Hamiltonian EVP is performed and found 
to yield a generalized EVP with a diagonal overlap matrix. 

As a first test and paradigmatic example the developed numerical approaches are applied to the Wannier-Mott exciton, 
which is solved numerically in the reciprocal space. Using realistic model parameters for the reduced effective mass 
and screening constant, the numerical convergence is systematically studied with special focus on the convergence with respect to the 
$\vk$-point sampling.
The well known analytical solution of a hydrogen-like exciton series is reproduced for the lowest bound pair states corresponding 
to the principle quantum numbers $n=1\dots 3$, including the expected degeneracies of exciton states and the oscillator strengths.

Further, the developed approaches are applied to the non-metallic materials MgO and InN, which both show Wannier-Mott-like excitons. 
The respective excitons, however, differ drastically in terms of their localization and consequently convergence, due to the 
very different electronic structures found for both materials. Nevertheless, the convergence trends identified for 
the Wannier-Mott model are found to hold and allow for the extrapolation of $\vk$-point converged binding energies 
for the strongest bound excitons.
If only the electronic screening is taken into account, the converged binding energies clearly exceed the experimental results.
This is attributed to the influence of the lattice polarization on the screening and therewith the exciton binding.

Further, the achieved level of convergence, together with the account for the complete electronic structure, 
allows for the discussion of effects, which are not addressable in $\vk\cdot\vec{p}$ theory. 
For InN, for instance, the often neglected $\vk$-dependence of the single-particle oscillator strengths
is shown to influence the polarization dependence of the lowest excitons drastically.

\section*{Acknowledgements}
We are grateful to Karsten Hannewald and J\"urgen Furthm\"uller for fruitful discussions, 
and acknowledge the financial support from the Deutsche
Forschungsgemeinschaft (Project No. Be1346/18-2 and Be1346/20-1), the European
Community in the framework of the NOE NANOQUANTA
(Contract No.~NMP4-CT-2004-500198), the ITN RAINBOW (GA No. 213238-2), and the Carl-Zeiss-Stiftung. 
Further, we thank the H{\"o}chstleistungs-Rechenzentrum M{\"u}nchen for its grants.


\begin{thebibliography}{75}
\expandafter\ifx\csname natexlab\endcsname\relax\def\natexlab#1{#1}\fi
\expandafter\ifx\csname bibnamefont\endcsname\relax
  \def\bibnamefont#1{#1}\fi
\expandafter\ifx\csname bibfnamefont\endcsname\relax
  \def\bibfnamefont#1{#1}\fi
\expandafter\ifx\csname citenamefont\endcsname\relax
  \def\citenamefont#1{#1}\fi
\expandafter\ifx\csname url\endcsname\relax
  \def\url#1{\texttt{#1}}\fi
\expandafter\ifx\csname urlprefix\endcsname\relax\def\urlprefix{URL }\fi
\providecommand{\bibinfo}[2]{#2}
\providecommand{\eprint}[2][]{\url{#2}}

\bibitem[{\citenamefont{Sham and Rice}(1966)}]{Sham:1966:MPD}
\bibinfo{author}{\bibfnamefont{L.~J.} \bibnamefont{Sham}} \bibnamefont{and}
  \bibinfo{author}{\bibfnamefont{T.~M.} \bibnamefont{Rice}},
  \bibinfo{journal}{Phys. Rev.} \textbf{\bibinfo{volume}{144}},
  \bibinfo{pages}{708} (\bibinfo{year}{1966}).

\bibitem[{\citenamefont{Hanke and Sham}(1979)}]{Hanke:1979:MPE}
\bibinfo{author}{\bibfnamefont{W.}~\bibnamefont{Hanke}} \bibnamefont{and}
  \bibinfo{author}{\bibfnamefont{L.~J.} \bibnamefont{Sham}},
  \bibinfo{journal}{Phys. Rev. Lett.} \textbf{\bibinfo{volume}{43}},
  \bibinfo{pages}{387} (\bibinfo{year}{1979}).

\bibitem[{\citenamefont{Strinati}(1988)}]{Strinati:1988:AGF}
\bibinfo{author}{\bibfnamefont{G.}~\bibnamefont{Strinati}},
  \bibinfo{journal}{Rivista del Nuovo Cimento} \textbf{\bibinfo{volume}{11}},
  \bibinfo{pages}{1} (\bibinfo{year}{1988}).

\bibitem[{\citenamefont{Onida et~al.}(2002)\citenamefont{Onida, Reining, and
  Rubio}}]{Onida:2002:EED}
\bibinfo{author}{\bibfnamefont{G.}~\bibnamefont{Onida}},
  \bibinfo{author}{\bibfnamefont{L.}~\bibnamefont{Reining}}, \bibnamefont{and}
  \bibinfo{author}{\bibfnamefont{A.}~\bibnamefont{Rubio}},
  \bibinfo{journal}{Rev. Mod. Phys.} \textbf{\bibinfo{volume}{74}},
  \bibinfo{pages}{601} (\bibinfo{year}{2002}).

\bibitem[{\citenamefont{Hedin}(1965)}]{Hedin:1965:F}
\bibinfo{author}{\bibfnamefont{L.}~\bibnamefont{Hedin}},
  \bibinfo{journal}{Phys. Rev.} \textbf{\bibinfo{volume}{139}},
  \bibinfo{pages}{A796} (\bibinfo{year}{1965}).

\bibitem[{\citenamefont{Hybertsen and Louie}(1985)}]{Hybertsen:louie:1985b}
\bibinfo{author}{\bibfnamefont{M.~S.} \bibnamefont{Hybertsen}}
  \bibnamefont{and} \bibinfo{author}{\bibfnamefont{S.~G.} \bibnamefont{Louie}},
  \bibinfo{journal}{Phys. Rev. B} \textbf{\bibinfo{volume}{32}},
  \bibinfo{pages}{7005} (\bibinfo{year}{1985}).

\bibitem[{\citenamefont{Godby et~al.}(1987)\citenamefont{Godby, Schl{\"u}ter,
  and Sham}}]{Godby:1987}
\bibinfo{author}{\bibfnamefont{R.~W.} \bibnamefont{Godby}},
  \bibinfo{author}{\bibfnamefont{M.}~\bibnamefont{Schl{\"u}ter}},
  \bibnamefont{and} \bibinfo{author}{\bibfnamefont{L.~J.} \bibnamefont{Sham}},
  \bibinfo{journal}{Phys. Rev. B} \textbf{\bibinfo{volume}{35}},
  \bibinfo{pages}{4170} (\bibinfo{year}{1987}).

\bibitem[{\citenamefont{Bechstedt}(1992)}]{Bechstedt:1992}
\bibinfo{author}{\bibfnamefont{F.}~\bibnamefont{Bechstedt}},
  \bibinfo{journal}{Adv. Solid State Phys.} \textbf{\bibinfo{volume}{32}},
  \bibinfo{pages}{161} (\bibinfo{year}{1992}).

\bibitem[{\citenamefont{Aryasetiawan and
  Gunnarsson}(1998)}]{Aryasetiawan:1998:GM}
\bibinfo{author}{\bibfnamefont{F.}~\bibnamefont{Aryasetiawan}}
  \bibnamefont{and}
  \bibinfo{author}{\bibfnamefont{O.}~\bibnamefont{Gunnarsson}},
  \bibinfo{journal}{Rep. Prog. Phys.} \textbf{\bibinfo{volume}{61}},
  \bibinfo{pages}{237} (\bibinfo{year}{1998}).

\bibitem[{\citenamefont{Aulbur et~al.}(2000)\citenamefont{Aulbur, J{\"o}nsson,
  and Wilkins}}]{Aulbur:2000:SSP}
\bibinfo{author}{\bibfnamefont{W.~G.} \bibnamefont{Aulbur}},
  \bibinfo{author}{\bibfnamefont{L.}~\bibnamefont{J{\"o}nsson}},
  \bibnamefont{and} \bibinfo{author}{\bibfnamefont{J.~W.}
  \bibnamefont{Wilkins}}, \emph{\bibinfo{title}{Solid State Physics: Advances
  in Research and Applications}} (\bibinfo{publisher}{Academic},
  \bibinfo{address}{San Diego}, \bibinfo{year}{2000}),
  vol.~\bibinfo{volume}{54}, chap. \bibinfo{chapter}{Quasiparticle calculations
  in solids}, p.~\bibinfo{pages}{1}.

\bibitem[{\citenamefont{Albrecht et~al.}(1998)\citenamefont{Albrecht, Reining,
  {Del Sole}, and Onida}}]{Albrecht:1998:ICE}
\bibinfo{author}{\bibfnamefont{S.}~\bibnamefont{Albrecht}},
  \bibinfo{author}{\bibfnamefont{L.}~\bibnamefont{Reining}},
  \bibinfo{author}{\bibfnamefont{R.}~\bibnamefont{{Del Sole}}},
  \bibnamefont{and} \bibinfo{author}{\bibfnamefont{G.}~\bibnamefont{Onida}},
  \bibinfo{journal}{Phys. Rev. Lett.} \textbf{\bibinfo{volume}{80}},
  \bibinfo{pages}{4510} (\bibinfo{year}{1998}).

\bibitem[{\citenamefont{Rohlfing and
  Louie}(1998{\natexlab{a}})}]{Rohlfing:1998:EHE}
\bibinfo{author}{\bibfnamefont{M.}~\bibnamefont{Rohlfing}} \bibnamefont{and}
  \bibinfo{author}{\bibfnamefont{S.~G.} \bibnamefont{Louie}},
  \bibinfo{journal}{Phys. Rev. Lett.} \textbf{\bibinfo{volume}{81}},
  \bibinfo{pages}{2312} (\bibinfo{year}{1998}{\natexlab{a}}).

\bibitem[{\citenamefont{Benedict et~al.}(1998)\citenamefont{Benedict, Shirley,
  and Bohn}}]{Benedict:1998:OAI}
\bibinfo{author}{\bibfnamefont{L.~X.} \bibnamefont{Benedict}},
  \bibinfo{author}{\bibfnamefont{E.~L.} \bibnamefont{Shirley}},
  \bibnamefont{and} \bibinfo{author}{\bibfnamefont{R.~B.} \bibnamefont{Bohn}},
  \bibinfo{journal}{Phys. Rev. Lett.} \textbf{\bibinfo{volume}{80}},
  \bibinfo{pages}{4514} (\bibinfo{year}{1998}).

\bibitem[{\citenamefont{Schmidt et~al.}(2003)\citenamefont{Schmidt, Glutsch,
  Hahn, and Bechstedt}}]{Schmidt:2003:EMS}
\bibinfo{author}{\bibfnamefont{W.~G.} \bibnamefont{Schmidt}},
  \bibinfo{author}{\bibfnamefont{S.}~\bibnamefont{Glutsch}},
  \bibinfo{author}{\bibfnamefont{P.~H.} \bibnamefont{Hahn}}, \bibnamefont{and}
  \bibinfo{author}{\bibfnamefont{F.}~\bibnamefont{Bechstedt}},
  \bibinfo{journal}{Phys. Rev. B} \textbf{\bibinfo{volume}{67}},
  \bibinfo{pages}{085307} (\bibinfo{year}{2003}).

\bibitem[{\citenamefont{Rohlfing and Louie}(1999)}]{Rohlfing:1999:EOS}
\bibinfo{author}{\bibfnamefont{M.}~\bibnamefont{Rohlfing}} \bibnamefont{and}
  \bibinfo{author}{\bibfnamefont{S.~G.} \bibnamefont{Louie}},
  \bibinfo{journal}{Phys. Rev. Lett.} \textbf{\bibinfo{volume}{83}},
  \bibinfo{pages}{856} (\bibinfo{year}{1999}).

\bibitem[{\citenamefont{Hahn et~al.}(2002)\citenamefont{Hahn, Schmidt, and
  Bechstedt}}]{Hahn:2002:BEE}
\bibinfo{author}{\bibfnamefont{P.~H.} \bibnamefont{Hahn}},
  \bibinfo{author}{\bibfnamefont{W.~G.} \bibnamefont{Schmidt}},
  \bibnamefont{and}
  \bibinfo{author}{\bibfnamefont{F.}~\bibnamefont{Bechstedt}},
  \bibinfo{journal}{Phys. Rev. Lett.} \textbf{\bibinfo{volume}{88}},
  \bibinfo{pages}{016402} (\bibinfo{year}{2002}).

\bibitem[{\citenamefont{Rohlfing and
  Louie}(1998{\natexlab{b}})}]{Rohlfing:1998:EEO}
\bibinfo{author}{\bibfnamefont{M.}~\bibnamefont{Rohlfing}} \bibnamefont{and}
  \bibinfo{author}{\bibfnamefont{S.~G.} \bibnamefont{Louie}},
  \bibinfo{journal}{Phys. Rev. Lett.} \textbf{\bibinfo{volume}{80}},
  \bibinfo{pages}{3320} (\bibinfo{year}{1998}{\natexlab{b}}).

\bibitem[{\citenamefont{Hahn et~al.}(2005{\natexlab{a}})\citenamefont{Hahn,
  Schmidt, and Bechstedt}}]{Hahn:2005:MEE}
\bibinfo{author}{\bibfnamefont{P.~H.} \bibnamefont{Hahn}},
  \bibinfo{author}{\bibfnamefont{W.~G.} \bibnamefont{Schmidt}},
  \bibnamefont{and}
  \bibinfo{author}{\bibfnamefont{F.}~\bibnamefont{Bechstedt}},
  \bibinfo{journal}{Phys. Rev. B} \textbf{\bibinfo{volume}{72}},
  \bibinfo{pages}{245425} (\bibinfo{year}{2005}{\natexlab{a}}).

\bibitem[{\citenamefont{Hahn et~al.}(2005{\natexlab{b}})\citenamefont{Hahn,
  Schmidt, Seino, Preuss, Bechstedt, and Bernholc}}]{Hahn:2005:OAW}
\bibinfo{author}{\bibfnamefont{P.~H.} \bibnamefont{Hahn}},
  \bibinfo{author}{\bibfnamefont{W.~G.} \bibnamefont{Schmidt}},
  \bibinfo{author}{\bibfnamefont{K.}~\bibnamefont{Seino}},
  \bibinfo{author}{\bibfnamefont{M.}~\bibnamefont{Preuss}},
  \bibinfo{author}{\bibfnamefont{F.}~\bibnamefont{Bechstedt}},
  \bibnamefont{and} \bibinfo{author}{\bibfnamefont{J.}~\bibnamefont{Bernholc}},
  \bibinfo{journal}{Phys. Rev. Lett.} \textbf{\bibinfo{volume}{94}},
  \bibinfo{pages}{037404} (\bibinfo{year}{2005}{\natexlab{b}}).

\bibitem[{\citenamefont{Garbuio et~al.}(2006)\citenamefont{Garbuio, Cascella,
  Reining, Sole, and Pulci}}]{Garbuio:2006}
\bibinfo{author}{\bibfnamefont{V.}~\bibnamefont{Garbuio}},
  \bibinfo{author}{\bibfnamefont{M.}~\bibnamefont{Cascella}},
  \bibinfo{author}{\bibfnamefont{L.}~\bibnamefont{Reining}},
  \bibinfo{author}{\bibfnamefont{R.~D.} \bibnamefont{Sole}}, \bibnamefont{and}
  \bibinfo{author}{\bibfnamefont{O.}~\bibnamefont{Pulci}},
  \bibinfo{journal}{Phys. Rev. Lett.} \textbf{\bibinfo{volume}{97}},
  \bibinfo{pages}{137402} (\bibinfo{year}{2006}).

\bibitem[{\citenamefont{Kittel}(1996)}]{Kittel:1996}
\bibinfo{author}{\bibfnamefont{C.}~\bibnamefont{Kittel}},
  \emph{\bibinfo{title}{Introduction to Solid State Physics}}
  (\bibinfo{publisher}{John Wiley and Sons}, \bibinfo{address}{New York,
  Chichester}, \bibinfo{year}{1996}).

\bibitem[{\citenamefont{Benedict and Shirley}(1999)}]{Benedict:1999:ICI}
\bibinfo{author}{\bibfnamefont{L.~X.} \bibnamefont{Benedict}} \bibnamefont{and}
  \bibinfo{author}{\bibfnamefont{E.~L.} \bibnamefont{Shirley}},
  \bibinfo{journal}{Phys. Rev. B} \textbf{\bibinfo{volume}{59}},
  \bibinfo{pages}{5441} (\bibinfo{year}{1999}).

\bibitem[{\citenamefont{Bechstedt et~al.}(2005)\citenamefont{Bechstedt, Seino,
  Hahn, and Schmidt}}]{Bechstedt:2005}
\bibinfo{author}{\bibfnamefont{F.}~\bibnamefont{Bechstedt}},
  \bibinfo{author}{\bibfnamefont{K.}~\bibnamefont{Seino}},
  \bibinfo{author}{\bibfnamefont{P.~H.} \bibnamefont{Hahn}}, \bibnamefont{and}
  \bibinfo{author}{\bibfnamefont{W.~G.} \bibnamefont{Schmidt}},
  \bibinfo{journal}{Phys. Rev. B} \textbf{\bibinfo{volume}{72}},
  \bibinfo{pages}{245425} (\bibinfo{year}{2005}).

\bibitem[{\citenamefont{Hahn et~al.}(2005{\natexlab{c}})\citenamefont{Hahn,
  Seino, Schmidt, Furthm{\"u}ller, and Bechstedt}}]{Hahn:2005}
\bibinfo{author}{\bibfnamefont{P.~H.} \bibnamefont{Hahn}},
  \bibinfo{author}{\bibfnamefont{K.}~\bibnamefont{Seino}},
  \bibinfo{author}{\bibfnamefont{W.~G.} \bibnamefont{Schmidt}},
  \bibinfo{author}{\bibfnamefont{J.}~\bibnamefont{Furthm{\"u}ller}},
  \bibnamefont{and}
  \bibinfo{author}{\bibfnamefont{F.}~\bibnamefont{Bechstedt}},
  \bibinfo{journal}{phys. stat. sol. (b)} \textbf{\bibinfo{volume}{242}},
  \bibinfo{pages}{2720} (\bibinfo{year}{2005}{\natexlab{c}}).

\bibitem[{\citenamefont{Yu and Cardona}(1999)}]{Yu:1999:FS}
\bibinfo{author}{\bibfnamefont{P.~Y.} \bibnamefont{Yu}} \bibnamefont{and}
  \bibinfo{author}{\bibfnamefont{M.}~\bibnamefont{Cardona}},
  \emph{\bibinfo{title}{Fundamentals of Semiconductors}}
  (\bibinfo{publisher}{Springer-Verlag}, \bibinfo{address}{Berlin},
  \bibinfo{year}{1999}).

\bibitem[{\citenamefont{Elliott}(1957)}]{Elliot:1957}
\bibinfo{author}{\bibfnamefont{R.~J.} \bibnamefont{Elliott}},
  \bibinfo{journal}{Phys. Rev. B} \textbf{\bibinfo{volume}{108}},
  \bibinfo{pages}{1384} (\bibinfo{year}{1957}).

\bibitem[{\citenamefont{Shinada and Sugano}(1966)}]{Shinada:1966}
\bibinfo{author}{\bibfnamefont{M.}~\bibnamefont{Shinada}} \bibnamefont{and}
  \bibinfo{author}{\bibfnamefont{S.}~\bibnamefont{Sugano}},
  \bibinfo{journal}{J. Phys. Soc. Japan} \textbf{\bibinfo{volume}{21}},
  \bibinfo{pages}{1936} (\bibinfo{year}{1966}).

\bibitem[{\citenamefont{Rohlfing and Louie}(2000)}]{Rohlfing:2000:EHE}
\bibinfo{author}{\bibfnamefont{M.}~\bibnamefont{Rohlfing}} \bibnamefont{and}
  \bibinfo{author}{\bibfnamefont{S.~G.} \bibnamefont{Louie}},
  \bibinfo{journal}{Phys. Rev. B} \textbf{\bibinfo{volume}{62}},
  \bibinfo{pages}{4927} (\bibinfo{year}{2000}).

\bibitem[{\citenamefont{Laskowski et~al.}(2005)\citenamefont{Laskowski,
  Christensen, Santi, and Ambrosch-Draxl}}]{Laskowski:2005}
\bibinfo{author}{\bibfnamefont{R.}~\bibnamefont{Laskowski}},
  \bibinfo{author}{\bibfnamefont{N.~E.} \bibnamefont{Christensen}},
  \bibinfo{author}{\bibfnamefont{G.}~\bibnamefont{Santi}}, \bibnamefont{and}
  \bibinfo{author}{\bibfnamefont{C.}~\bibnamefont{Ambrosch-Draxl}},
  \bibinfo{journal}{Phys. Rev. B} \textbf{\bibinfo{volume}{72}},
  \bibinfo{pages}{035204} (\bibinfo{year}{2005}).

\bibitem[{\citenamefont{Laskowski and Christensen}(2006)}]{Laskowski:2006}
\bibinfo{author}{\bibfnamefont{R.}~\bibnamefont{Laskowski}} \bibnamefont{and}
  \bibinfo{author}{\bibfnamefont{N.~E.} \bibnamefont{Christensen}},
  \bibinfo{journal}{Phys. Rev. B} \textbf{\bibinfo{volume}{73}},
  \bibinfo{pages}{045201} (\bibinfo{year}{2006}).

\bibitem[{\citenamefont{Denisov and Makarov}(1973)}]{Denisov:1973}
\bibinfo{author}{\bibfnamefont{M.}~\bibnamefont{Denisov}} \bibnamefont{and}
  \bibinfo{author}{\bibfnamefont{V.}~\bibnamefont{Makarov}},
  \bibinfo{journal}{phys. stat. sol. (b)} \textbf{\bibinfo{volume}{56}},
  \bibinfo{pages}{9} (\bibinfo{year}{1973}).

\bibitem[{\citenamefont{Fu et~al.}(1999)\citenamefont{Fu, Wang, and
  Zunger}}]{Fu:1999}
\bibinfo{author}{\bibfnamefont{H.}~\bibnamefont{Fu}},
  \bibinfo{author}{\bibfnamefont{L.-W.} \bibnamefont{Wang}}, \bibnamefont{and}
  \bibinfo{author}{\bibfnamefont{A.}~\bibnamefont{Zunger}},
  \bibinfo{journal}{Phys. Rev. B} \textbf{\bibinfo{volume}{59}},
  \bibinfo{pages}{5568} (\bibinfo{year}{1999}).

\bibitem[{\citenamefont{R{\"o}dl et~al.}(2008)\citenamefont{R{\"o}dl, Fuchs,
  Furthm{\"u}ller, and Bechstedt}}]{Roedl:2008}
\bibinfo{author}{\bibfnamefont{C.}~\bibnamefont{R{\"o}dl}},
  \bibinfo{author}{\bibfnamefont{F.}~\bibnamefont{Fuchs}},
  \bibinfo{author}{\bibfnamefont{J.}~\bibnamefont{Furthm{\"u}ller}},
  \bibnamefont{and}
  \bibinfo{author}{\bibfnamefont{F.}~\bibnamefont{Bechstedt}},
  \bibinfo{journal}{Phys. Rev. B}  (\bibinfo{year}{2008}),
  \bibinfo{note}{accepted}.

\bibitem[{\citenamefont{Anisimov et~al.}(1991)\citenamefont{Anisimov, Zaanen,
  and Andersen}}]{Anisimov:1991}
\bibinfo{author}{\bibfnamefont{V.~I.} \bibnamefont{Anisimov}},
  \bibinfo{author}{\bibfnamefont{J.}~\bibnamefont{Zaanen}}, \bibnamefont{and}
  \bibinfo{author}{\bibfnamefont{O.~K.} \bibnamefont{Andersen}},
  \bibinfo{journal}{Phys. Rev. B} \textbf{\bibinfo{volume}{44}},
  \bibinfo{pages}{943} (\bibinfo{year}{1991}).

\bibitem[{\citenamefont{Seidl et~al.}(1996)\citenamefont{Seidl, G{\"o}rling,
  Vogl, Majewski, and Levy}}]{Seidl:1996}
\bibinfo{author}{\bibfnamefont{A.}~\bibnamefont{Seidl}},
  \bibinfo{author}{\bibfnamefont{A.}~\bibnamefont{G{\"o}rling}},
  \bibinfo{author}{\bibfnamefont{P.}~\bibnamefont{Vogl}},
  \bibinfo{author}{\bibfnamefont{J.~A.} \bibnamefont{Majewski}},
  \bibnamefont{and} \bibinfo{author}{\bibfnamefont{M.}~\bibnamefont{Levy}},
  \bibinfo{journal}{Phys. Rev. B} \textbf{\bibinfo{volume}{53}},
  \bibinfo{pages}{3764} (\bibinfo{year}{1996}).

\bibitem[{\citenamefont{Fuchs et~al.}(2007)\citenamefont{Fuchs,
  Furthm{\"u}ller, Bechstedt, Shishkin, and Kresse}}]{Fuchs:2007:PRB}
\bibinfo{author}{\bibfnamefont{F.}~\bibnamefont{Fuchs}},
  \bibinfo{author}{\bibfnamefont{J.}~\bibnamefont{Furthm{\"u}ller}},
  \bibinfo{author}{\bibfnamefont{F.}~\bibnamefont{Bechstedt}},
  \bibinfo{author}{\bibfnamefont{M.}~\bibnamefont{Shishkin}}, \bibnamefont{and}
  \bibinfo{author}{\bibfnamefont{G.}~\bibnamefont{Kresse}},
  \bibinfo{journal}{Phys. Rev. B} \textbf{\bibinfo{volume}{76}},
  \bibinfo{pages}{115109} (\bibinfo{year}{2007}).

\bibitem[{\citenamefont{Puschnig and Ambrosch-Draxl}(2002)}]{Puschnig:2002}
\bibinfo{author}{\bibfnamefont{P.}~\bibnamefont{Puschnig}} \bibnamefont{and}
  \bibinfo{author}{\bibfnamefont{C.}~\bibnamefont{Ambrosch-Draxl}},
  \bibinfo{journal}{Phys. Rev. B} \textbf{\bibinfo{volume}{66}},
  \bibinfo{pages}{165105} (\bibinfo{year}{2002}).

\bibitem[{\citenamefont{Schleife et~al.}(2007)\citenamefont{Schleife, R{\"o}dl,
  Fuchs, Furthm{\"u}ller, and Bechstedt}}]{Schleife:2007:exciton}
\bibinfo{author}{\bibfnamefont{A.}~\bibnamefont{Schleife}},
  \bibinfo{author}{\bibfnamefont{C.}~\bibnamefont{R{\"o}dl}},
  \bibinfo{author}{\bibfnamefont{F.}~\bibnamefont{Fuchs}},
  \bibinfo{author}{\bibfnamefont{J.}~\bibnamefont{Furthm{\"u}ller}},
  \bibnamefont{and}
  \bibinfo{author}{\bibfnamefont{F.}~\bibnamefont{Bechstedt}},
  \bibinfo{journal}{Appl. Phys. Lett.} \textbf{\bibinfo{volume}{91}},
  \bibinfo{pages}{241915} (\bibinfo{year}{2007}).

\bibitem[{\citenamefont{Kalkreuter and Simma}(1996)}]{Kalkreuter:1995}
\bibinfo{author}{\bibfnamefont{T.}~\bibnamefont{Kalkreuter}} \bibnamefont{and}
  \bibinfo{author}{\bibfnamefont{H.}~\bibnamefont{Simma}},
  \bibinfo{journal}{Comput. Phys. Commun.} \textbf{\bibinfo{volume}{93}},
  \bibinfo{pages}{33} (\bibinfo{year}{1996}).

\bibitem[{LAP()}]{LAPACK}
\bibinfo{howpublished}{Linear Algebra PACKage, http://www.netlib.org/lapack/}.

\bibitem[{\citenamefont{Monkhorst and Pack}(1976)}]{Monkhorst:1976}
\bibinfo{author}{\bibfnamefont{H.~J.} \bibnamefont{Monkhorst}}
  \bibnamefont{and} \bibinfo{author}{\bibfnamefont{J.~D.} \bibnamefont{Pack}},
  \bibinfo{journal}{Phys. Rev. B} \textbf{\bibinfo{volume}{13}},
  \bibinfo{pages}{5188} (\bibinfo{year}{1976}).

\bibitem[{\citenamefont{Kresse and Furthm{\"u}ller}(1996)}]{Kresse:1996:EIT}
\bibinfo{author}{\bibfnamefont{G.}~\bibnamefont{Kresse}} \bibnamefont{and}
  \bibinfo{author}{\bibfnamefont{J.}~\bibnamefont{Furthm{\"u}ller}},
  \bibinfo{journal}{Comp. Mat. Sci.} \textbf{\bibinfo{volume}{6}},
  \bibinfo{pages}{15} (\bibinfo{year}{1996}).

\bibitem[{\citenamefont{Shishkin and Kresse}(2006)}]{Shishkin:2006}
\bibinfo{author}{\bibfnamefont{M.}~\bibnamefont{Shishkin}} \bibnamefont{and}
  \bibinfo{author}{\bibfnamefont{G.}~\bibnamefont{Kresse}},
  \bibinfo{journal}{Phys. Rev. B} \textbf{\bibinfo{volume}{74}},
  \bibinfo{pages}{035101} (\bibinfo{year}{2006}).

\bibitem[{\citenamefont{Bl{\"o}chl}(1994)}]{Blochl:1994:PAW}
\bibinfo{author}{\bibfnamefont{P.~E.} \bibnamefont{Bl{\"o}chl}},
  \bibinfo{journal}{Phys. Rev. B} \textbf{\bibinfo{volume}{50}},
  \bibinfo{pages}{17953} (\bibinfo{year}{1994}).

\bibitem[{\citenamefont{Kresse and Joubert}(1999)}]{Kresse:1998:UPP}
\bibinfo{author}{\bibfnamefont{G.}~\bibnamefont{Kresse}} \bibnamefont{and}
  \bibinfo{author}{\bibfnamefont{D.}~\bibnamefont{Joubert}},
  \bibinfo{journal}{Phys. Rev. B} \textbf{\bibinfo{volume}{59}},
  \bibinfo{pages}{1758} (\bibinfo{year}{1999}).

\bibitem[{\citenamefont{Bechstedt et~al.}(1992)\citenamefont{Bechstedt, {Del
  Sole}, Cappellini, and Reining}}]{Bechstedt:1992:EMC}
\bibinfo{author}{\bibfnamefont{F.}~\bibnamefont{Bechstedt}},
  \bibinfo{author}{\bibfnamefont{R.}~\bibnamefont{{Del Sole}}},
  \bibinfo{author}{\bibfnamefont{G.}~\bibnamefont{Cappellini}},
  \bibnamefont{and} \bibinfo{author}{\bibfnamefont{L.}~\bibnamefont{Reining}},
  \bibinfo{journal}{Solid State Commun.} \textbf{\bibinfo{volume}{84}},
  \bibinfo{pages}{765} (\bibinfo{year}{1992}).

\bibitem[{\citenamefont{Wannier}(1937)}]{Wannier:1937}
\bibinfo{author}{\bibfnamefont{G.~H.} \bibnamefont{Wannier}},
  \bibinfo{journal}{Phys. Rev.} \textbf{\bibinfo{volume}{52}},
  \bibinfo{pages}{191} (\bibinfo{year}{1937}).

\bibitem[{\citenamefont{Mott}(1938)}]{Mott:1938}
\bibinfo{author}{\bibfnamefont{N.~F.} \bibnamefont{Mott}},
  \bibinfo{journal}{Trans. Faraday Soc.} \textbf{\bibinfo{volume}{34}},
  \bibinfo{pages}{500} (\bibinfo{year}{1938}).

\bibitem[{\citenamefont{Roessler and Walker}(1967)}]{Roessler:1967}
\bibinfo{author}{\bibfnamefont{D.~M.} \bibnamefont{Roessler}} \bibnamefont{and}
  \bibinfo{author}{\bibfnamefont{W.~C.} \bibnamefont{Walker}},
  \bibinfo{journal}{Phys. Rev.} \textbf{\bibinfo{volume}{159}},
  \bibinfo{pages}{733} (\bibinfo{year}{1967}).

\bibitem[{\citenamefont{Schleife et~al.}(2006)\citenamefont{Schleife, Fuchs,
  Furthm{\"u}ller, and Bechstedt}}]{Schleife:2006}
\bibinfo{author}{\bibfnamefont{A.}~\bibnamefont{Schleife}},
  \bibinfo{author}{\bibfnamefont{F.}~\bibnamefont{Fuchs}},
  \bibinfo{author}{\bibfnamefont{J.}~\bibnamefont{Furthm{\"u}ller}},
  \bibnamefont{and}
  \bibinfo{author}{\bibfnamefont{F.}~\bibnamefont{Bechstedt}},
  \bibinfo{journal}{Phys. Rev. B} \textbf{\bibinfo{volume}{73}},
  \bibinfo{pages}{245212} (\bibinfo{year}{2006}).

\bibitem[{\citenamefont{van Schilfgaarde et~al.}(2006)\citenamefont{van
  Schilfgaarde, Kotani, and Faleev}}]{vanSchilfgaarde:2006}
\bibinfo{author}{\bibfnamefont{M.}~\bibnamefont{van Schilfgaarde}},
  \bibinfo{author}{\bibfnamefont{T.}~\bibnamefont{Kotani}}, \bibnamefont{and}
  \bibinfo{author}{\bibfnamefont{S.}~\bibnamefont{Faleev}},
  \bibinfo{journal}{Phys. Rev. Lett.} \textbf{\bibinfo{volume}{96}},
  \bibinfo{pages}{226402} (\bibinfo{year}{2006}).

\bibitem[{\citenamefont{Shishkin et~al.}(2007)\citenamefont{Shishkin, Marsman,
  and Kresse}}]{Shishkin:2007}
\bibinfo{author}{\bibfnamefont{M.}~\bibnamefont{Shishkin}},
  \bibinfo{author}{\bibfnamefont{M.}~\bibnamefont{Marsman}}, \bibnamefont{and}
  \bibinfo{author}{\bibfnamefont{G.}~\bibnamefont{Kresse}},
  \bibinfo{journal}{Phys. Rev. B} \textbf{\bibinfo{volume}{99}},
  \bibinfo{pages}{246403} (\bibinfo{year}{2007}).

\bibitem[{\citenamefont{Baldereschi and Lipari}(1971)}]{Baldereschi:1971}
\bibinfo{author}{\bibfnamefont{A.}~\bibnamefont{Baldereschi}} \bibnamefont{and}
  \bibinfo{author}{\bibfnamefont{N.~C.} \bibnamefont{Lipari}},
  \bibinfo{journal}{Phys. Rev. B} \textbf{\bibinfo{volume}{3}},
  \bibinfo{pages}{439} (\bibinfo{year}{1971}).

\bibitem[{\citenamefont{Martienssen and Warlimont}(2005)}]{Martienssen:2005}
\bibinfo{editor}{\bibfnamefont{W.}~\bibnamefont{Martienssen}} \bibnamefont{and}
  \bibinfo{editor}{\bibfnamefont{H.}~\bibnamefont{Warlimont}}, eds.,
  \emph{\bibinfo{title}{Springer Handbook of Condensed Matter and Materials
  Data}} (\bibinfo{publisher}{Springer-Verlag}, \bibinfo{address}{Berlin},
  \bibinfo{year}{2005}), p. \bibinfo{pages}{660}.

\bibitem[{\citenamefont{Whited et~al.}(1973)\citenamefont{Whited, Flaten, and
  Walker}}]{whited73}
\bibinfo{author}{\bibfnamefont{R.~C.} \bibnamefont{Whited}},
  \bibinfo{author}{\bibfnamefont{C.~J.} \bibnamefont{Flaten}},
  \bibnamefont{and} \bibinfo{author}{\bibfnamefont{W.~C.}
  \bibnamefont{Walker}}, \bibinfo{journal}{Solid State Commun.}
  \textbf{\bibinfo{volume}{13}}, \bibinfo{pages}{1903} (\bibinfo{year}{1973}).

\bibitem[{\citenamefont{Shindo}(1970)}]{Shindo:1970}
\bibinfo{author}{\bibfnamefont{K.}~\bibnamefont{Shindo}}, \bibinfo{journal}{J.
  Phys. Soc. Jpn.} \textbf{\bibinfo{volume}{29}}, \bibinfo{pages}{287}
  (\bibinfo{year}{1970}).

\bibitem[{\citenamefont{Zimmermann}(1971)}]{Zimmermann:1971}
\bibinfo{author}{\bibfnamefont{R.}~\bibnamefont{Zimmermann}},
  \bibinfo{journal}{phys. stat. sol. (b)} \textbf{\bibinfo{volume}{48}},
  \bibinfo{pages}{603} (\bibinfo{year}{1971}).

\bibitem[{\citenamefont{Whited and Walker}(1969)}]{whited69}
\bibinfo{author}{\bibfnamefont{R.~C.} \bibnamefont{Whited}} \bibnamefont{and}
  \bibinfo{author}{\bibfnamefont{W.~C.} \bibnamefont{Walker}},
  \bibinfo{journal}{Phys. Rev. Lett.} \textbf{\bibinfo{volume}{22}},
  \bibinfo{pages}{1428} (\bibinfo{year}{1969}).

\bibitem[{\citenamefont{Bechstedt et~al.}(1980)\citenamefont{Bechstedt,
  Enderlein, and Koch}}]{Bechstedt:1980}
\bibinfo{author}{\bibfnamefont{F.}~\bibnamefont{Bechstedt}},
  \bibinfo{author}{\bibfnamefont{R.}~\bibnamefont{Enderlein}},
  \bibnamefont{and} \bibinfo{author}{\bibfnamefont{M.}~\bibnamefont{Koch}},
  \bibinfo{journal}{phys. stat. sol. (b)} \textbf{\bibinfo{volume}{99}},
  \bibinfo{pages}{61} (\bibinfo{year}{1980}).

\bibitem[{\citenamefont{Wang et~al.}(2004)\citenamefont{Wang, Rohlfing,
  Kr{\"u}ger, and Pollmann}}]{Wang:2004}
\bibinfo{author}{\bibfnamefont{N.}~\bibnamefont{Wang}},
  \bibinfo{author}{\bibfnamefont{M.}~\bibnamefont{Rohlfing}},
  \bibinfo{author}{\bibfnamefont{P.}~\bibnamefont{Kr{\"u}ger}},
  \bibnamefont{and} \bibinfo{author}{\bibfnamefont{J.}~\bibnamefont{Pollmann}},
  \bibinfo{journal}{Appl. Phys. A} \textbf{\bibinfo{volume}{78}},
  \bibinfo{pages}{213} (\bibinfo{year}{2004}).

\bibitem[{\citenamefont{Davydov et~al.}(2002)\citenamefont{Davydov, Klochikhin,
  Emtsev, Ivanov, Vekshin, Bechstedt, Furthm{\"u}ller, Harima, Mudryi,
  Hashimoto et~al.}}]{Davydov:2002}
\bibinfo{author}{\bibfnamefont{V.~Y.} \bibnamefont{Davydov}},
  \bibinfo{author}{\bibfnamefont{A.~A.} \bibnamefont{Klochikhin}},
  \bibinfo{author}{\bibfnamefont{V.~V.} \bibnamefont{Emtsev}},
  \bibinfo{author}{\bibfnamefont{S.~V.} \bibnamefont{Ivanov}},
  \bibinfo{author}{\bibfnamefont{V.~V.} \bibnamefont{Vekshin}},
  \bibinfo{author}{\bibfnamefont{F.}~\bibnamefont{Bechstedt}},
  \bibinfo{author}{\bibfnamefont{J.}~\bibnamefont{Furthm{\"u}ller}},
  \bibinfo{author}{\bibfnamefont{H.}~\bibnamefont{Harima}},
  \bibinfo{author}{\bibfnamefont{A.~V.} \bibnamefont{Mudryi}},
  \bibinfo{author}{\bibfnamefont{A.}~\bibnamefont{Hashimoto}},
  \bibnamefont{et~al.}, \bibinfo{journal}{phys. stat. sol (b)} p.
  \bibinfo{pages}{230} (\bibinfo{year}{2002}).

\bibitem[{\citenamefont{Matsuoka et~al.}(2002)\citenamefont{Matsuoka, Okamoto,
  Nakao, Harima, and Kurimoto}}]{Matsuoka:2002}
\bibinfo{author}{\bibfnamefont{T.}~\bibnamefont{Matsuoka}},
  \bibinfo{author}{\bibfnamefont{H.}~\bibnamefont{Okamoto}},
  \bibinfo{author}{\bibfnamefont{M.}~\bibnamefont{Nakao}},
  \bibinfo{author}{\bibfnamefont{H.}~\bibnamefont{Harima}}, \bibnamefont{and}
  \bibinfo{author}{\bibfnamefont{E.}~\bibnamefont{Kurimoto}},
  \bibinfo{journal}{Appl. Phys. Lett.} \textbf{\bibinfo{volume}{81}},
  \bibinfo{pages}{1246} (\bibinfo{year}{2002}).

\bibitem[{\citenamefont{Wu et~al.}(2002)\citenamefont{Wu, Walukiewicz, Yu,
  Ager~III, Haller, Lu, Schaff, Saito, and Nanishi}}]{Wu:2002}
\bibinfo{author}{\bibfnamefont{J.}~\bibnamefont{Wu}},
  \bibinfo{author}{\bibfnamefont{W.}~\bibnamefont{Walukiewicz}},
  \bibinfo{author}{\bibfnamefont{K.}~\bibnamefont{Yu}},
  \bibinfo{author}{\bibfnamefont{J.}~\bibnamefont{Ager~III}},
  \bibinfo{author}{\bibfnamefont{E.}~\bibnamefont{Haller}},
  \bibinfo{author}{\bibfnamefont{H.}~\bibnamefont{Lu}},
  \bibinfo{author}{\bibfnamefont{W.~J.} \bibnamefont{Schaff}},
  \bibinfo{author}{\bibfnamefont{Y.}~\bibnamefont{Saito}}, \bibnamefont{and}
  \bibinfo{author}{\bibfnamefont{Y.}~\bibnamefont{Nanishi}},
  \bibinfo{journal}{Appl. Phys. Lett.} \textbf{\bibinfo{volume}{80}},
  \bibinfo{pages}{3967} (\bibinfo{year}{2002}).

\bibitem[{\citenamefont{Furthm{\"u}ller
  et~al.}(2005)\citenamefont{Furthm{\"u}ller, Hahn, Fuchs, and
  Bechstedt}}]{Furthmuller:2005}
\bibinfo{author}{\bibfnamefont{J.}~\bibnamefont{Furthm{\"u}ller}},
  \bibinfo{author}{\bibfnamefont{P.~H.} \bibnamefont{Hahn}},
  \bibinfo{author}{\bibfnamefont{F.}~\bibnamefont{Fuchs}}, \bibnamefont{and}
  \bibinfo{author}{\bibfnamefont{F.}~\bibnamefont{Bechstedt}},
  \bibinfo{journal}{Phys. Rev. B} \textbf{\bibinfo{volume}{72}},
  \bibinfo{pages}{205106} (\bibinfo{year}{2005}).

\bibitem[{\citenamefont{Rinke et~al.}(2006)\citenamefont{Rinke, Scheffler,
  Qteish, Winkelnkemper, and Bimberg}}]{Rinke:2006}
\bibinfo{author}{\bibfnamefont{P.}~\bibnamefont{Rinke}},
  \bibinfo{author}{\bibfnamefont{M.}~\bibnamefont{Scheffler}},
  \bibinfo{author}{\bibfnamefont{A.}~\bibnamefont{Qteish}},
  \bibinfo{author}{\bibfnamefont{M.}~\bibnamefont{Winkelnkemper}},
  \bibnamefont{and} \bibinfo{author}{\bibfnamefont{D.}~\bibnamefont{Bimberg}},
  \bibinfo{journal}{Appl. Phys. Lett.} \textbf{\bibinfo{volume}{89}},
  \bibinfo{pages}{161919} (\bibinfo{year}{2006}).

\bibitem[{\citenamefont{Heyd et~al.}(2003)\citenamefont{Heyd, Scuseria, and
  Ernzerhof}}]{Heyd:2003}
\bibinfo{author}{\bibfnamefont{J.}~\bibnamefont{Heyd}},
  \bibinfo{author}{\bibfnamefont{G.~E.} \bibnamefont{Scuseria}},
  \bibnamefont{and}
  \bibinfo{author}{\bibfnamefont{M.}~\bibnamefont{Ernzerhof}},
  \bibinfo{journal}{J. Chem. Phys.} \textbf{\bibinfo{volume}{118}},
  \bibinfo{pages}{8207} (\bibinfo{year}{2003}).

\bibitem[{\citenamefont{Anisimov et~al.}(1993)\citenamefont{Anisimov, Solovyev,
  Korotin, Czy\.{z}yk, and Sawatzky}}]{Anisimov:1993}
\bibinfo{author}{\bibfnamefont{V.~I.} \bibnamefont{Anisimov}},
  \bibinfo{author}{\bibfnamefont{I.~V.} \bibnamefont{Solovyev}},
  \bibinfo{author}{\bibfnamefont{M.~A.} \bibnamefont{Korotin}},
  \bibinfo{author}{\bibfnamefont{M.~T.} \bibnamefont{Czy\.{z}yk}},
  \bibnamefont{and} \bibinfo{author}{\bibfnamefont{G.~A.}
  \bibnamefont{Sawatzky}}, \bibinfo{journal}{Phys. Rev. B}
  \textbf{\bibinfo{volume}{48}}, \bibinfo{pages}{16929} (\bibinfo{year}{1993}).

\bibitem[{\citenamefont{Franchini et~al.}(2005)\citenamefont{Franchini, Bayer,
  Podloucky, Paier, and Kresse}}]{Franchini:2005}
\bibinfo{author}{\bibfnamefont{C.}~\bibnamefont{Franchini}},
  \bibinfo{author}{\bibfnamefont{V.}~\bibnamefont{Bayer}},
  \bibinfo{author}{\bibfnamefont{R.}~\bibnamefont{Podloucky}},
  \bibinfo{author}{\bibfnamefont{J.}~\bibnamefont{Paier}}, \bibnamefont{and}
  \bibinfo{author}{\bibfnamefont{G.}~\bibnamefont{Kresse}},
  \bibinfo{journal}{Phys. Rev. B} \textbf{\bibinfo{volume}{72}},
  \bibinfo{pages}{045132} (\bibinfo{year}{2005}).

\bibitem[{\citenamefont{Goldhahn et~al.}(2006)\citenamefont{Goldhahn, Schley,
  Winzer, Rakel, Cobet, Esser, Lu, and Schaff}}]{Goldhahn:2006}
\bibinfo{author}{\bibfnamefont{R.}~\bibnamefont{Goldhahn}},
  \bibinfo{author}{\bibfnamefont{P.}~\bibnamefont{Schley}},
  \bibinfo{author}{\bibfnamefont{A.~T.} \bibnamefont{Winzer}},
  \bibinfo{author}{\bibfnamefont{M.}~\bibnamefont{Rakel}},
  \bibinfo{author}{\bibfnamefont{C.}~\bibnamefont{Cobet}},
  \bibinfo{author}{\bibfnamefont{N.}~\bibnamefont{Esser}},
  \bibinfo{author}{\bibfnamefont{H.}~\bibnamefont{Lu}}, \bibnamefont{and}
  \bibinfo{author}{\bibfnamefont{W.~J.} \bibnamefont{Schaff}},
  \bibinfo{journal}{J. Cryst. Growth} \textbf{\bibinfo{volume}{288}},
  \bibinfo{pages}{273} (\bibinfo{year}{2006}).

\bibitem[{\citenamefont{Fuchs et~al.}()\citenamefont{Fuchs, Furthm{\"u}ller,
  and Bechstedt}}]{Fuchs:2008:unpub}
\bibinfo{author}{\bibfnamefont{F.}~\bibnamefont{Fuchs}},
  \bibinfo{author}{\bibfnamefont{J.}~\bibnamefont{Furthm{\"u}ller}},
  \bibnamefont{and}
  \bibinfo{author}{\bibfnamefont{F.}~\bibnamefont{Bechstedt}},
  \bibinfo{note}{unpublished}.

\bibitem[{\citenamefont{Schley et~al.}(2007)\citenamefont{Schley, Goldhahn,
  Winzer, Gobsch, Cimalla, Ambacher, Lu, Schaff, Kurouchi, Nanishi
  et~al.}}]{Schley:2007}
\bibinfo{author}{\bibfnamefont{P.}~\bibnamefont{Schley}},
  \bibinfo{author}{\bibfnamefont{R.}~\bibnamefont{Goldhahn}},
  \bibinfo{author}{\bibfnamefont{A.~T.} \bibnamefont{Winzer}},
  \bibinfo{author}{\bibfnamefont{G.}~\bibnamefont{Gobsch}},
  \bibinfo{author}{\bibfnamefont{V.}~\bibnamefont{Cimalla}},
  \bibinfo{author}{\bibfnamefont{O.}~\bibnamefont{Ambacher}},
  \bibinfo{author}{\bibfnamefont{H.}~\bibnamefont{Lu}},
  \bibinfo{author}{\bibfnamefont{W.~J.} \bibnamefont{Schaff}},
  \bibinfo{author}{\bibfnamefont{M.}~\bibnamefont{Kurouchi}},
  \bibinfo{author}{\bibfnamefont{Y.}~\bibnamefont{Nanishi}},
  \bibnamefont{et~al.}, \bibinfo{journal}{Phys. Rev. B}
  \textbf{\bibinfo{volume}{75}}, \bibinfo{pages}{205204}
  (\bibinfo{year}{2007}).

\bibitem[{\citenamefont{Goldhahn et~al.}(2004)\citenamefont{Goldhahn, Winzer,
  Cimalla, Ambacher, Cobet, Richter, Esser, Furthm{\"u}ller, Bechstedt, Lu
  et~al.}}]{Goldhahn:2004}
\bibinfo{author}{\bibfnamefont{R.}~\bibnamefont{Goldhahn}},
  \bibinfo{author}{\bibfnamefont{A.~T.} \bibnamefont{Winzer}},
  \bibinfo{author}{\bibfnamefont{V.}~\bibnamefont{Cimalla}},
  \bibinfo{author}{\bibfnamefont{O.}~\bibnamefont{Ambacher}},
  \bibinfo{author}{\bibfnamefont{C.}~\bibnamefont{Cobet}},
  \bibinfo{author}{\bibfnamefont{W.}~\bibnamefont{Richter}},
  \bibinfo{author}{\bibfnamefont{N.}~\bibnamefont{Esser}},
  \bibinfo{author}{\bibfnamefont{J.}~\bibnamefont{Furthm{\"u}ller}},
  \bibinfo{author}{\bibfnamefont{F.}~\bibnamefont{Bechstedt}},
  \bibinfo{author}{\bibfnamefont{H.}~\bibnamefont{Lu}}, \bibnamefont{et~al.},
  \bibinfo{journal}{Superlatt. Microstruct.} \textbf{\bibinfo{volume}{36}},
  \bibinfo{pages}{591} (\bibinfo{year}{2004}).

\bibitem[{\citenamefont{Chuang and Chang}(1996)}]{Chuang:1996}
\bibinfo{author}{\bibfnamefont{S.~L.} \bibnamefont{Chuang}} \bibnamefont{and}
  \bibinfo{author}{\bibfnamefont{C.~S.} \bibnamefont{Chang}},
  \bibinfo{journal}{Phys. Rev.} \textbf{\bibinfo{volume}{54}},
  \bibinfo{pages}{2491} (\bibinfo{year}{1996}).

\bibitem[{\citenamefont{Thomas and Hopfield}(1959)}]{Thomas:1959}
\bibinfo{author}{\bibfnamefont{D.~G.} \bibnamefont{Thomas}} \bibnamefont{and}
  \bibinfo{author}{\bibfnamefont{J.~J.} \bibnamefont{Hopfield}},
  \bibinfo{journal}{Phys. Rev.} \textbf{\bibinfo{volume}{116}},
  \bibinfo{pages}{573} (\bibinfo{year}{1959}).

\bibitem[{\citenamefont{Davydov et~al.}(1999)\citenamefont{Davydov, Klochikhin,
  Smirnov, Emtsev, Petrikov, Abroyan, Titov, Goncharuk, Smirnov, Mamutin
  et~al.}}]{Davydov:1999}
\bibinfo{author}{\bibfnamefont{V.}~\bibnamefont{Davydov}},
  \bibinfo{author}{\bibfnamefont{A.}~\bibnamefont{Klochikhin}},
  \bibinfo{author}{\bibfnamefont{M.}~\bibnamefont{Smirnov}},
  \bibinfo{author}{\bibfnamefont{V.}~\bibnamefont{Emtsev}},
  \bibinfo{author}{\bibfnamefont{V.}~\bibnamefont{Petrikov}},
  \bibinfo{author}{\bibfnamefont{I.}~\bibnamefont{Abroyan}},
  \bibinfo{author}{\bibfnamefont{A.}~\bibnamefont{Titov}},
  \bibinfo{author}{\bibfnamefont{I.}~\bibnamefont{Goncharuk}},
  \bibinfo{author}{\bibfnamefont{A.}~\bibnamefont{Smirnov}},
  \bibinfo{author}{\bibfnamefont{V.}~\bibnamefont{Mamutin}},
  \bibnamefont{et~al.}, \bibinfo{journal}{phys. stat. sol. (b)}
  \textbf{\bibinfo{volume}{216}}, \bibinfo{pages}{779} (\bibinfo{year}{1999}).

\end{thebibliography}

\end{document}